\documentclass{article}
\usepackage[utf8]{inputenc}	% For diacritical symbols (tildes, umlaute, etc.)
\usepackage[T2A]{fontenc} % For text in Russian
\usepackage{graphicx} % Required for inserting images
\usepackage{setspace} % Double or other spacing
\usepackage{amsmath, amsfonts}  % Math fonts
\usepackage{microtype}  % Nice format
\usepackage{natbib}  % Bibliography
\usepackage[toc,page]{appendix}  % To have several appendices
\usepackage{authblk} % Nicer way of indicating affiliations
\usepackage{xurl} % Sentence breaks for URLs
\usepackage{booktabs}  % Nicer tables

\usepackage{subcaption}  % Side by side graphs in figures
\usepackage{chngcntr}  % Change numbering of figures/tables in section

 \textwidth15.75cm \evensidemargin5mm \oddsidemargin5mm
\title{The economic consequences of geopolitical fragmentation: Evidence from the Cold War\footnote{We thank Rodrigo Peña for excellent research assistance and Marina Diakonova for her help in locating the digitized copies of the statistical reports on foreign trade of the USSR. We thank Evi Pappa for helpful comments and suggestions. Heid gratefully acknowledges financial support from the Spanish Ministry of Universities (Ministerio de Universidades, Plan de Recuperación, Transformación y Resiliencia) under María Zambrano contract MAZ/2021/04(UP2021-021) financed by the European Union - NextGenerationEU, from PID2020-114646RB-C42 funded by MCIN-AEI/10.13039/501100011033 (a project from the Ministerio de Ciencia e Innovación (MCIN), Agencia Estatal de Investigación (AEI), Spain), from Prometeo project CIPROM/2022/50 financed by the Generalitat Valenciana, and from the Plan de Promoción de la Investigación de la Universitat Jaume I, UJI-B2022-36-(22I587). The funding sources had no involvement in the writing of this paper nor the decision to submit it for publication. Declarations of interest: none. The views expressed in this paper are those of the authors and do therefore not necessarily reflect those of the Banco de Espa\~na or the Eurosystem. }}

\author[1]{Rodolfo G. Campos}
\author[2]{Benedikt Heid}
\author[1]{Jacopo Timini}
\affil[1]{Banco de España}
\affil[2]{Universitat Jaume I, Grup d'investigaci\'{o} en Integraci\'{o} Econ\`{o}mica (INTECO), University of Adelaide, CESifo}

\date{\today}

\begin{document}

\maketitle
\begin{abstract} \noindent
The Cold War was the defining episode of geopolitical fragmentation in the twentieth century. Trade between East and West across the Iron Curtain (a symbolical and physical barrier dividing Europe into two distinct areas) was restricted, but the severity of these restrictions varied over time. We quantify the trade and welfare effects of the Iron Curtain and show how the difficulty of trading across the Iron Curtain fluctuated throughout the Cold War. Using a novel dataset on trade between the two economic blocs and a quantitative trade model, we find that while the Iron Curtain at its height represented a tariff equivalent of 48\% in 1951, trade between East and West gradually became easier until the fall of the Berlin Wall in 1989. Despite the easing of trade restrictions, we estimate that the Iron Curtain roughly halved East-West trade flows and caused substantial welfare losses in the Eastern bloc countries that persisted until the end of the Cold War. Conversely, the Iron Curtain led to an increase in intra-bloc trade, especially in the Eastern bloc, which outpaced the integration of Western Europe in the run-up to the formation of the European Union.
\end{abstract}
% Word count: 4 words

\textbf{Keywords}: International trade, Cold War, Iron Curtain, geopolitical fragmentation, trade costs of borders

\textbf{JEL codes}: F13, F14, N74

\clearpage

\begin{doublespace}

\section{Introduction}

The Iron Curtain represented the political, military, and cultural barriers that existed between the East and West blocs throughout the Cold War era, represented most notoriously by the Berlin Wall.\footnote{The term ``Iron Curtain'' was coined by Winston Churchill in his ``Sinews of Peace'' speech in 1946 to describe the division between the Soviet-dominated East and the Western democracies.} The Iron Curtain was therefore both a physical boundary and a complex system of policies of trade restrictions that impeded the flow of goods and services. However, whereas the physical border did not move, the political restrictions were not set in stone. Contrary to common perceptions of the Cold War as a period of absolute isolation between East and West, trade policies toward each other varied considerably over time, and little is known about the extent to which these changes affected trade barriers. 

%Parallels with today's situation come to mind: 
As the world grapples with signs of increasing geopolitical fragmentation, it is important to reflect on the lessons of the past. Understanding the impact of trade barriers on countries separated by the Iron Curtain can provide insight into the potential consequences of future fragmentation. By quantifying historical trade barriers, policymakers and scholars can better anticipate the economic and political outcomes of current episodes of geopolitical fragmentation.

Our paper delivers such a quantification. Particularly, we estimate the level of trade barriers imposed by the Iron Curtain, and how they fluctuated over time. In a second step, we use our estimates and a state-of-the art quantitative trade model that belongs to the class of ``universal gravity'' models described by \citet{AllenArkolakisTakahashi:20} to simulate the trade and welfare effects of a counterfactual world without the Iron Curtain. A major challenge is the lack of complete historical data on bilateral trade flows for important countries belonging to the Eastern bloc. The IMF's DOTS database, for example, which is one of the main sources of bilateral trade data in the postwar period, does not include trade flows involving East Germany and the USSR, either as exporters or importers, for many years.

We overcome this problem by collecting information from several editions of the statistical yearbooks of East Germany and the statistical reviews of foreign trade of the Soviet Union. We use exactly the same methodology as the IMF to incorporate these additional observations into the DOTS database. Where the DOTS database contains data for specific years, we cross-check our approach to ensure that our methodology closely matches the reported information.

With this completed dataset, we use a structural gravity model to estimate the effect of the Iron Curtain to bilateral international trade flows. We identify this effect by comparing international trade between countries across the Iron Curtain, i.e., across the Eastern and Western bloc, with international trade within the two blocs. 

\textbf{Results}. Our estimates suggest a progressive easing of trade barriers. While there were intermittent fluctuations in the difficulty of trading across the Iron Curtain, the overall pattern shows a gradual easing of trade restrictions over time. Our method allows us to express the trade barriers between East and West as a tariff-equivalent. At its height, the Iron Curtain had an effect similar to a 48\% ad valorem tariff. The tariff-equivalent measure declined sharply in the second half of the 1950s and throughout the 1960s, settling at around 25\% in the 1970s and 1980s.

Increased trade barriers between the blocs were the direct result of the Iron Curtain. Over time, however, the two economic blocs became more integrated within themselves, so trade between countries within the blocs became easier. Our estimates show the increasing easing of trade within the Western bloc countries in the period from the signing of the Treaties of Rome in 1957, which marked the beginning of the institutional process that led to the European Union, to the period immediately before the creation of the single market in 1993. The Eastern bloc established the Council for Mutual Economic Assistance (COMECON) in 1949, which also led to deeper integration. Interestingly, and perhaps surprisingly, our estimates suggest that integration within the Eastern bloc outpaced integration in the West. This movement toward greater integration continued steadily through the end of the Cold War, while Western integration stalled in the 1970s.

Not all countries in Europe were part of the Western or Eastern bloc. Switzerland, Ireland, and Sweden were not aligned with either bloc, but can be considered closer to the West in terms of their economic relations, while Austria and Finland were considered to be closer to neutral during the Cold War. Yugoslavia was originally aligned with the East, but moved to a more neutral position. We examine how trade barriers evolved between these three groups of countries and the blocs on either side of the Iron Curtain. Western-leaning countries faced growing trade barriers with the East, comparable to those in neutral countries like Austria and Finland, at least until the early 1970s. Yugoslavia, on the other hand, experienced increased trade costs with both blocs. This implies that the Eastern bloc countries were the most disconnected from the rest of the world, as they experienced increased trade costs not only with the West, but also with non-aligned or neutral countries.

We perform counterfactual simulations based on general equilibrium trade model that belongs to the class of ``universal gravity'' models described by \citet{AllenArkolakisTakahashi:20}. We find that the Iron Curtain roughly halved trade flows between East and West. Welfare losses in the median Eastern country were close to 1\% of per capita consumption per year. In particular, welfare losses increase toward the end of the period, reflecting the increasing burden imposed on the East by its economic isolation from the rest of the world prior to the end of the Cold War. This suggests that the lack of international trade may have been behind the drive to liberalize East bloc economies at the end of the Cold War, with initiatives such as perestroika, which was launched by Gorbachev in the Soviet Union.

\textbf{Related literature}. Our paper quantifies the effect of the Iron Curtain on trade using historical trade data during and after the period of the Cold War, demonstrating how it varies over time. 
Previous research that has examined a constant effect of the Iron Curtain on trade, such as the studies by \citet{vanBergeijk:15} and \citet{EggerFoellmiSchetterTorun:23} use cross-sectional data for a single year. \citet{BeestermöllerRauch:18} also use a gravity model to analyze the Iron Curtain, but focus on the period \textit{after} its fall and on trade relations among countries formerly part of the defunct Austro-Hungarian Monarchy; their main finding is that these countries tended to trade more after the fall of the Iron Curtain than implied by standard gravity models, suggesting a persistent trade-enhancing institutional legacy. The focus on the post-Cold War period is also the focus of \citet{NitschWolf:13}, who analyze the evolution of intra-German trade flows for the period after the fall of the Berlin Wall and find persistent differences in trade patterns along the former East-West German border. Other papers on the topic of the Berlin Wall include \citet{AhlfeldtReddingSturmWolf:15}, who use the Berlin Wall as a source of exogenous variation to explain changes in urban structure, and  \citet{ReddingSturm:08}, who exploit the division and subsequent reunification of Germany to explain population dynamics across the country. Other papers that analyze trade integration between Eastern and Western Europe after the fall of the Berlin Wall include \citet{Piazolo:97}, \citet{JakabKovacsOszlay:01}, \citet{BussièreFidrmucSchnatz:08}, and \citet{RavishankarStack:14}.

Our paper is also related to the literature that attempts to estimate the effects of borders on trade. This literature begins with the work of \citet{McCallum:95}, who compares regional trade within and between Canadian and US regions, and the modern treatment by \citet{AndersonvanWincoop:03}.\footnote{Related to this work, \citet{AgnostevaAndersonYotov:19} also use trade between Canadian provinces to examine the variation in domestic trade costs across geographic units.} Many recent studies have examined the effects of borders on trade around the world. For example, \citet{LawlessNearyStudnicka:19} focus on trade between Ireland and Northern Ireland, while \citet{SantamariaVenturaYeşilbayraktar:23}, \citet{FrenschFidrmucRindler:23}, and \citet{Spornberger:22} look more broadly at borders within Europe. \citet{CarterPoast:20} and \citet{KamwelaDemenavanBergeijk:23} instead examine the effect of border walls and argue that physical border barriers significantly reduce bilateral trade.\footnote{In a related vein, \citet{LarchTanYotov:23} show how the measurement of the impact of borders on trade can be used to study the ex ante effects of trade liberalization and protection.} In the context of the literature of border effects, our study adds to the body of knowledge by quantifying the impact over time of one of the most important borders in recent European history.

Finally, our paper also contributes to the literature that examines the relationship between geopolitics and trade. This body of research examines how geopolitical factors affect international trade. \citet{MartinThoenigMayer:08} and \citet{KarlssonHedberg:21} study how civil and interstate wars affect bilateral trade and \citet{BergerEasterlyNunnSatyanath:13} analyze how political influence resulting from CIA interventions increases U.S. exports to targeted countries.  \citet{Yu:10}, \citet{FuchsKlann:13}, \citet{DuJuRamirezYao:17}, \citet{FelbermayrSyropoulosYalcinYotov:20}, \citet{LarchShikherSyropoulosYotov:22}, \citet{FlachHeilandLarchSteiningerTeti:23}, and \citet{JäkelØstervigYalcini:23} examine how bilateral political relations or broader political institutions (such as democracy or sanctions) affect trade. An emerging strand of this literature analyzes how so-called ``trade fragmentation''---a multipolar world characterized by the emergence of multiple regional economic blocs---may affect international trade patterns and economic dynamics. Recent examples in this vein include \citet{AiyarChenEbekeGarcia-SaltosGudmundssonIlyinaKangurKunaratskulRodriguezRutaSchulzeSoderbergTrevino:23}, \citet{AiyarPresbiteroRuta:23}, and \citet{CamposEstefania-FloresFurceriTimini:23}. The Iron Curtain serves as an important case study in this context, as it represents one of the most salient instances of geopolitical fragmentation in recent history.

\textbf{Overview}. The rest of the paper is structured as follows. Section~\ref{sec:historical_context} gives a brief account of the historical context. Section~\ref{sec:methodology} presents our methodology and Section~\ref{sec:data} describes our data. We relegate the details of how we obtained and cross-checked the data for East Germany and the USSR to the appendix. In Section~\ref{sec:results} we present the estimates of trade barriers between blocs and additional empirical results, while in Section~\ref{sec:quantification} we quantify the costs of the Iron Curtain in terms of trade flows and welfare using counterfactual simulations. Section~\ref{sec:conclusion} concludes.

\section{Historical context and the need for a quantification of trade barriers} \label{sec:historical_context}

During the Cold War, trade and foreign policy were closely intertwined, especially in shaping relations between the Eastern and Western blocs. As a result, political and ideological differences between the communist-dominated Eastern bloc and the capitalist-dominated Western bloc resulted in significant restrictions on trade between the blocs.

On the one side of the Iron Curtain, the Eastern bloc, led by the Soviet Union, adopted a command economy with centralized planning and state ownership of industry. This system lacked market mechanisms and therefore discouraged foreign trade because the state was directly involved in foreign trade decisions. Existing trade relations were driven by political or ideological motives. In addition, the Eastern bloc had limited convertibility, meaning that their currencies could not be freely exchanged for hard (or Western) currencies. This made transactions difficult and hindered trade between the blocs. The Soviet Union and its allies relied on bilateral agreements and barter trade rather than convertible currencies.

On the other side of the Iron Curtain, the Western bloc established a set of trade policy instruments linked to its foreign policy goal of containing the spread of communism. These included a wide range of instruments such as export controls, particularly the export of strategic materials and technological goods under the auspices of the Coordinating Committee for Multilateral Export Controls (COCOM); sanctions; import barriers (such as quotas or licensing requirements); and restrictions on the provision of trade finance.

In parallel, both blocs sought to coordinate trade within their borders. The Eastern bloc countries formed the Council for Mutual Economic Assistance, an economic organization designed to coordinate economic planning and trade among the socialist states. Similarly, the Western bloc established several initiatives to facilitate trade and economic cooperation among its member countries: for example, the Organization for European Economic Cooperation (OEEC), which later evolved into the Organization for Economic Cooperation and Development (OECD); the European Coal and Steel Community (ECSC, later the European Economic Community, EEC, and the European Union, EU); or the European Free Trade Association (EFTA).

Moreover, while the Cold War led to an overall major divide in global trade, as both blocs restricted and limited trade with each other, the extent of these trade barriers and the degree to which they were enforced also varied over time, along with changes in the foreign policy landscape or reverberations of geopolitical tensions.

The many complex instruments used to either hinder or encourage trade between countries, both within and across trade blocs, make it difficult to obtain an accurate measure of the magnitude of trade barriers during this period, as is often the case when tariffs coexist with non-tariff measures. In this paper, we quantify the barriers to trade between the two blocs.

\section{Methodology} \label{sec:methodology}

For estimation purposes, we define the Western bloc as the group of the following European countries: Belgium, Denmark, France, Iceland, Italy, Luxembourg, Netherlands, Norway, Portugal, United Kingdom, Greece, Spain, West Germany. Similarly, we define the Eastern bloc as the group consisting of Bulgaria, Czechoslovakia, East Germany, Hungary, Poland, Romania, and the Soviet Union. For countries that separated after the fall of the Berlin Wall, we include all new countries in the group they were originally in. The Czech Republic and Slovakia are defined as part of the Eastern bloc, as are Russia, Estonia, Latvia, Lithuania, Belarus, Ukraine, and Moldova. Post-unification Germany is included in the Western bloc to reflect the relatively larger size of West Germany. We classify trade flows as crossing the Iron Curtain if they are from a country in the Western bloc to a country in the Eastern bloc or vice versa.

We estimate a state-of-the-art structural gravity model.\footnote{See \cite{HeadMayer:14} and \cite{YotovPiermartiniMonteiroLarch:16} for a review of current best practices.} We include exporter-year and importer-year dummy variables that account for so-called multilateral resistance terms, i.e., unobserved importer- and exporter-weighted averages of trade costs, to avoid omitted variable bias and thus mitigate endogeneity concerns raised by economic theory, see \cite{AndersonvanWincoop:03}.  

We follow standard practice and estimate the model using Poisson pseudo maximum likelihood (PPML), as originally proposed by \citet{SantosSilvaTenreyro:06}. This method provides consistent parameter estimates and trade cost elasticities in the presence of zero trade flows and heteroskedasticity. Moreover, it is the only estimator that is consistent with general equilibrium addition constraints, as shown by \citet{Fally:15}. Again, following standard practice, we use nominal trade data for all our estimations, as recommended by \citet{BaldwinTalgioni:07}. Note that the included importer-year and exporter-year fixed effects control for inflation differentials across countries.

\textbf{Baseline specification}. To identify the effect of interest, i.e., the impact of the Iron Curtain on international trade, we define a dummy variable (denoted $IC_{ij}$) that indicates whether trade flows that cross an international border do so from a country $i$ on one side of the Iron Curtain to a country $j$ on the other side. The equation we estimate is as follows: 
\begin{equation} \label{eq:specification}
    X_{ijt} = \exp(\gamma_t b_{ij} + \theta_t IC_{ij} + \phi_{it} + \psi_{jt} + \mathbf{z}_{ij}'\boldsymbol{\beta} + \varepsilon_{ijt}),
\end{equation}
where $X_{ijt}$ represents bilateral trade flows between exporter $i$ and importer $j$ in year $t$.  We identify trade flows that cross international borders, i.e., when the exporter $i$ is not the same country as the importer $j$, by $b_{ij}$, the so-called border dummy. This variable controls for the general effect of globalization in our specification, by allowing for a time-varying international border effect, following the recommendation of \cite{Yotov:12} and \cite{BergstrandLarchYotov:15}. $\phi_{it}$ and $\psi_{jt}$ are exporter-time and importer-time fixed effects They control for factors that vary at the country-year level, such as multilateral resistance terms. Note that they also control for the potential influence of variables related to a country's size, such as total sales or population. The vector $\mathbf{z}_{ij}$ is a vector of gravity variables (distance, common language, contiguity, and colonial relationship), and $\varepsilon_{ijt}$ is the error term. 

The parameters $\gamma_t$ measure the semi-elasticity of bilateral trade flows with respect to the presence of an international border for each year in the estimation sample. The coefficient of interest, $\theta_t$, measures the marginal contribution of the Iron Curtain to this elasticity for each year in the sample. To ensure a consistent estimation of the border effect, see \citet{HeidLarchYotov:21} and \citet{Yotov:22}, our data set includes both both international and domestic trade flows. This also allows to control for the potential trade diversion from international to domestic trade caused by the Iron Curtain, see \cite{DaiYotovZylkin:14}. However, as a robustness check, in Figure~\ref{fig:robustness_domestic} in Appendix~\ref{sec:appendix_empirical} we also show results from an estimation that does not use domestic trade flows. 

\textbf{Tariff equivalent measure of the Iron Curtain}. When presenting results, we first transform the estimated coefficients into their tariff equivalent. The formula is:
\begin{equation} \label{eq:marginal_effects}
    \text{Tariff equivalent}_t = 100 \times \left[\exp\left(-\frac{\hat{\theta}_t}{\epsilon}\right) - 1\right],
\end{equation}
where the notation $\hat{\theta}_t$ refers to the point estimates obtained, one for each year in the sample, and $\epsilon$ is the trade elasticity. Following \citet{HeadMayer:14}, we assume a value of $\epsilon = 5.03$ for the trade elasticity, although in Appendix~\ref{sec:appendix_empirical} (Figure~\ref{fig:raw_estimates}) we also report the coefficients from the estimation that do not depend on the value of the trade elasticity. 

The transformation converts the additional trade costs attributable to the Iron Curtain into the effect that an additional tariff would have if that tariff were imposed only on trade flows crossing the Iron Curtain. By construction, the possible values of the tariff equivalent of the Iron Curtain can be positive, zero, or even negative. A positive value indicates that the Iron Curtain restricts trade more than the typical national border. A value of zero means that crossing the Iron Curtain is no different than crossing a national border, while negative values are similar to a subsidy. In the latter case, the Iron Curtain would actually benefit trade.

\textbf{Caveats}. The border effects we identify serve as an indirect measure of the full set of policies implemented during the period under study. This approach has the advantage of capturing the aggregate effects of different policies that are difficult to quantify individually. However, this flexibility comes with the disadvantage of not being able to eliminate concerns that some of the variation is influenced by factors unrelated to government policies that we would prefer to exclude. Thus, a reasonable understanding of the summary measure of border effects, as implicitly applied in this study, is that it represents trade barriers during the Iron Curtain era, regardless of whether they were ultimately caused by government policies or not.

To guard against certain types of confounding factors, our specification includes multilateral resistance terms consisting of exporter-time and importer-time fixed effects. For confounding factors to significantly affect the estimation, they would have to vary at the bilateral level rather than at the country level. These confounding factors must not be constant over time, as they would already be accounted for by the standard gravity variables that vary bilaterally and are included in our specification. %We also conducted an estimation using bilateral fixed effects instead of the standard gravity variables and found that the effect of the Iron Curtain followed the same pattern as in our baseline specification. This suggests that the gravity variables adequately capture the time-invariant bilateral factors. 
Thus, for extraneous factors to affect the Iron Curtain border measure, they must vary bilaterally, and change over time.

\textbf{Alternative measures of trade costs used in the literature}. Our approach builds on the aggregate trade cost measure developed by \citet{JacksMeissnerNovy:08}, \citet{JacksMeissnerNovy:10}, and \citet{JacksMeissnerNovy:11} to analyze globalization over long periods of time, but with an important distinction relevant to our research question. Both approaches use the same underlying trade models to convert data into trade costs, but the difference lies in how border effects are accounted for. In our approach, we distinguish global border effects from those that are specific to the Iron Curtain. In addition, because our border effects are based on an estimation rather than a calibration process, we are able to conduct inference and provide measures of the precision of the estimated border effects using confidence intervals.

\textbf{Structural gravity models as a description of non-market economies}. A potential objection to using structural gravity models to analyze international trade flows of non-market economies might be that the countries of the Eastern bloc were not market economies, but centrally planned economies. Indeed, most descriptions of the gravity model derive the structural gravity equation for perfectly or imperfectly competitive markets populated by profit-maximizing firms, see, e.g., \citet{HeadMayer:14, HeidStälher:24}. However, the equilibrium trade flows implied by the structural gravity model of a market economy in its commonly used form are identical to the trade flows chosen by a central social planner.\footnote{In fact, in the canonical \citet{Armington:69} type of structural gravity popularized by \citet{AndersonvanWincoop:03}, only a representative household maximizes its utility, so the consumer's problem is identical to the social planner's problem.} The equivalence between the decentralized market equilibrium trade flows and the trade flows chosen by a central social planner holds even when firms are heterogeneous in their productivity. The key assumptions for these results are that demand has a constant elasticity of substitution (CES) and that the tradable goods produced are efficiently allocated across trading countries, i.e., there is no waste. Both are standard assumptions in international trade, and they underlie all structural gravity models. The intuition for this perhaps surprising result is that the structural gravity model describes the static efficient allocation of trade flows across countries for a given level of production, independent of the particular details of how production is actually organized. This is what \citet{Anderson:11} calls modularity.

This still leaves the empirical question of whether centrally planned economies were actually allocatively efficient. This is particularly relevant because the conventional wisdom is that socialist economies were notoriously inefficient. This perception stems from a conflation of static efficiency for a given level of technology, i.e. allocative efficiency, and dynamic efficiency, i.e. efficiency in terms of investment and innovation decisions.\footnote{For a lucid discussion of these issues, see \citet{Whitesell:90}.} The available empirical evidence on allocative efficiency finds that Soviet trade followed patterns of comparative advantage and was generally consistent with opportunity cost principles, see \citet{Rosefielde:74} for the case of Soviet international trade, and 
\citet{Murrell:91} for a broader review of the relevant empirical literature. As \citet{Murrell:91} notes, empirical approaches derived directly from neoclassical models---such as the structural gravity model---may well be applied to both centrally planned and market economies.\footnote{To be precise, \citet{Murrell:91}, p.\ 72 states: ``If one takes the neoclassical paradigm seriously in formulating empirical work, then one finds little to distinguish the two sets of economies.''} Finally, note that while Soviet economies were less productive than Western economies, such country-specific differences in absolute levels of technology are adequately captured by the fixed effects included in our structural gravity model.\footnote{See \citet{EatonKortum:02} for a derivation of a structural gravity model with absolute technology differences across countries.}

The upshot of the preceding discussion is that the structural gravity model is an adequate tool for analyzing trade flows in both market and centrally planned economies.

\section{Data} \label{sec:data}

\textbf{Bilateral trade data and domestic trade}. Our data on trade flows come from version 4 of the TRADHIST database \citep{FouquinHugot:16} and from statistical reports of East Germany and the Soviet Union. The TRADHIST database compiles historical bilateral trade flows of goods from various sources. Specifically for our period of interest, most of the data in the TRADHIST database comes from the IMF's DOTS database. Trade flows are measured in gross and nominal terms and are expressed in British pounds. To calculate domestic trade flows, we subtract nominal total exports from nominal GDP, both from the TRADHIST database. Although it would be consistent with economic theory to use gross output instead of GDP, there are no internationally comparable sources of gross output for our period of analysis. Moreover, recent studies \citep{CamposTiminiVidal:21} show that the presence of country and time fixed effects in gravity equations makes the distinction between GDP and gross output less important in practical applications. We use the variables distance, common language, contiguity and colonial relationship from the same database. The observations are annual. Due to the anomalous trade data during World War II and the limited availability of data prior to 1948, we start our estimation period in 1948. We extend the estimation period beyond the end of the Cold War and include the years 1991-2000.

\textbf{Additional data for East Germany and the USSR}. To overcome the lack of bilateral trade flows for East Germany and the Soviet Union in the original TRADHIST database, we have extracted trade data from historical reports. We followed the same steps to process the data as the IMF does for the DOTS database. To convert values from local currency to US dollars, we used exchange rates reported in the International Financial Statistics (IFS). Our methodology also included adjustments for c.i.f./f.o.b. discrepancies, so that imports are valued at c.i.f. value. 

We always prefer to use the value of a bilateral trade flow as reported by the importing country. Consequently, we replace observations in the TRADHIST database with the data we collect for all trade flows that are imports from East Germany or the USSR. We do this only when an observation in TRADHIST is either missing or zero and our collected data shows positive flows.

In the case of East Germany, data are collected from various editions of the \textit{Statistisches Jahrbuch der Deutschen Demokratischen Republik}. For the USSR, data are collected from the annual statistical reviews of foreign trade (Внешняя торговля СССР (Статистический обзор)). In Appendix~A we explain in detail the steps we followed to collect and process the data and also show that the bilateral trade values obtained in this way closely match those reported in DOTS for the relevant years (see Figures~\ref{fig:validation_1} through~\ref{fig:validation_6}). 

\textbf{Summary statistics}. Our final dataset contains 1,215,302 observations of bilateral trade flows, including domestic flows. As is common with bilateral trade data, a large fraction of these trade flows (49.8\%) are zero. Trade flows across the Iron Curtain account for 0.9\% of the observations.

\begin{table}[htbp]
    \caption{Summary statistics}
    \label{tab:summary_statistics}
    \centering
    \begin{tabular}{lcccc}
        \toprule
        Variable & Mean & Standard deviation &
        Minimum & Maximum \\ \midrule
        Bilateral trade (billion pounds sterling) & 0.27 & 18.85 & 0.00 & 6253.94\\
        Exporter belongs to the East bloc & 0.05 & 0.22 & 0.00 & 1.00 \\
        Importer belongs to the East bloc & 0.05 & 0.21 & 0.00 & 1.00 \\
        Exporter belongs to the West bloc & 0.10 & 0.30 & 0.00 & 1.00 \\
        Importer belongs to the West bloc & 0.10 & 0.30 & 0.00 & 1.00 \\
        Border dummy & 0.99 & 0.08 & 0.00 & 1.00 \\
        Log-distance & 8.66 & 1.04 & 0.00 & 9.90 \\
        Common language & 0.18 & 0.38 & 0.00 & 1.00 \\
        Contiguous & 0.02 & 0.14 & 0.00 & 1.00 \\
        Ever in colonial relationship & 0.02 & 0.12 & 0.00 & 1.00 \\
        \bottomrule
    \end{tabular}
    \vspace{3mm}
    
    \begin{minipage}{0.9\textwidth}
        \footnotesize \textbf{Notes}: Summary statistics for all variables are computed over 1,215,302 bilateral observations. With the exception of trade value and log distance, all other variables are dummy variables.
    \end{minipage}
\end{table}

Eastern bloc countries make up 5\% of exporting and importing countries. Western bloc countries account for 10\% of exporting and importing countries. Of all bilateral pairs, 18\% belong to countries that share a common language, while the share of bilateral pairs where the exporter and importer are contiguous countries or where both are part of a colonial relationship is 2\% in both cases.

%\clearpage
\section{Results} \label{sec:results}
\begin{figure}[htbp]
\begin{center}
\includegraphics[width=0.9\textwidth]{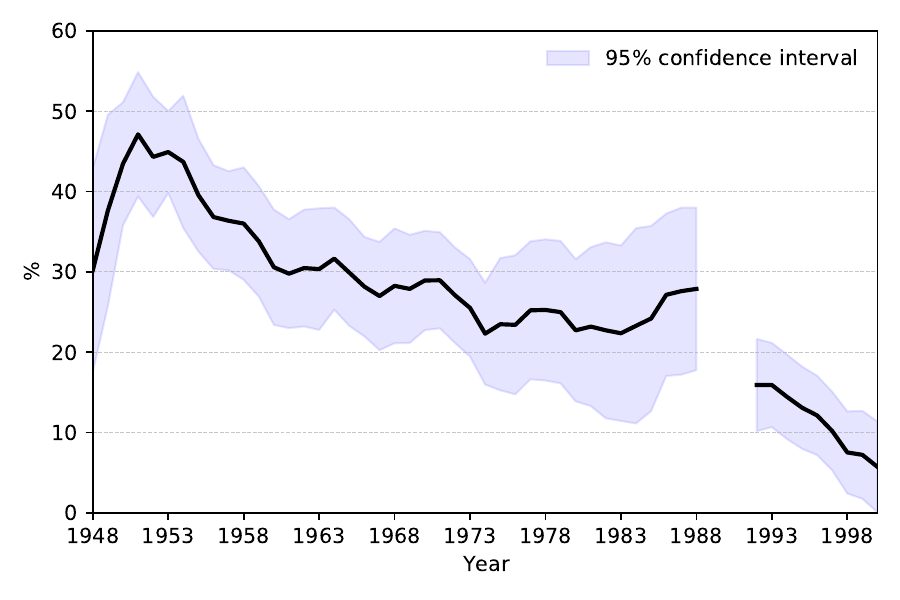}
\caption{Estimated tariff equivalent of the Iron Curtain\label{fig:IC1}}
\vspace{11pt}
\begin{minipage}{0.9\textwidth}
{\footnotesize 
\textbf{Notes}: The figure shows the estimated tariff equivalent of the Iron Curtain's borders measured in percentage points. The estimation uses the specification in~\eqref{eq:specification}. The tariff equivalent is calculated from the estimates $\hat{\theta}_t$ using the transformation $100\times[\exp(-\hat{\theta}_t/5.03)-1]$. The 95\% confidence interval is calculated using the delta method.}
\end{minipage}
\end{center}
\end{figure}

\textbf{The trade costs of the Iron Curtain}. The estimates from our baseline estimation are shown in Figure~\ref{fig:IC1}. They provide evidence of significant trade costs imposed by the Iron Curtain. Before the fall of the Berlin Wall in 1989, the Iron Curtain served as a barrier that imposed high trade costs between the Eastern and Western blocs. The chart also shows a dramatic shift in trade costs following the events of 1989. With the fall of the Iron Curtain, trade costs between the Eastern and Western blocs appear to have fallen rapidly. 

Focusing on the period of the Cold War, trade barriers between East and West showed a trend of progressive easing. While there were minor fluctuations in the difficulty of trading across the Iron Curtain, the overall pattern indicates a gradual reduction in trade restrictions. This easing of barriers between East and West can be translated into a tariff-equivalent measure.

At its height, the Iron Curtain imposed trade barriers equivalent to a 48\% ad valorem tariff. Over time, however, there was a decline in this tariff-equivalent measure, which started in the second half of the 1950s and continued throughout the 1960s. By the 1970s and 1980s, trade barriers had settled at around 25\%, indicating a significant reduction in trade barriers between East and West.

We formally test whether trade barriers associated with the Iron Curtain decreased over time by comparing four decade averages for the years 1948--1957, 1958--1967, 1968--1977, and 1978--1986. The statistical test for equality between the first two decades clearly rejects the null hypothesis (the p-value is 0.0004). The second and third decades are also statistically different (the p-value is 0.0081). For the last two decades, however, the null hypothesis of equality cannot be rejected at standard significance levels (the p-value is 0.6293).

\textbf{Discussion}. To put the ad valorem tariff into perspective, it is useful to compare them with other estimates available in the literature derived using similar models. Using virtually identical values for $\epsilon$, \citet{HeadMayer:21} indicate that the reduction in trade barriers between members of the European Union after the 1992 Maastricht Treaty deepened the Single Market was equivalent to an ad valorem tariff reduction of between 20\% and 30\% for trade in goods and between 10\% and 20\% for trade in services. Calculating the ad valorem tariff equivalent using the trade elasticity $\epsilon$=5.03, as we do, \citet{FelbermayrLarchYalcinYotov:24} results indicate that GATT/WTO membership is equivalent to an ad valorem tariff reduction of between 6\% and 13\%.

Overall, these numbers suggest that the estimated ad valorem tariff equivalents of major trade policy changes (such as entry into the GATT/WTO or the European Union) are substantially lower--by about two-thirds at most--than those observed during the Cold War. \citet{GlickTaylor:10} instead analyze the impact of wars on trade between belligerent parties, and find that the effects are comparable to an ad valorem tariff equivalent of about 40\%. This implies that our estimates suggest that during the height of the Cold War, the restrictions on trade between opposing factions were indeed similar to those experienced during actual wars.

%\clearpage

\textbf{Dissecting the effect of the Iron Curtain}.  In addition to the evolution of trade costs due to the presence of the Iron Curtain, we can also examine the variation of trade costs within the Eastern and Western bloc countries. To estimate this effect of trade costs, we extend the baseline specification in the following way:
\begin{equation} \label{eq:specification2}
    X_{ijt} = \exp(\gamma_t b_{ij} + \theta_t^{EW} EW_{ij} + \theta_t^{WE} WE_{ij} + \theta_t^{EE} EE_{ij} + \theta_t^{WW} WW_{ij} + \phi_{it} + \psi_{jt} + \mathbf{z}_{ij}'\boldsymbol{\beta}) + \varepsilon_{ijt}).
\end{equation}

In this equation, the dummy variables $EW$ and $WE$ correspond to international trade going from the Eastern bloc to the Western bloc and vice versa. Together, these two variables represent the Iron Curtain. We also include the variables $EE$ and $WW$, which indicate international trade within the Eastern and Western blocs, respectively. Note that these new variables are set to zero for domestic trade, so they only measure the impact on international trade flows.  Formally, the general formula for a variable of the form $AB$, where $A$ and $B$ are any two regions, is
\begin{equation}
AB_{ij} =
    \begin{cases}
        1 & \text{if } i \in A,\, j \in B,\, i \neq j,  \\
        0 & \text{otherwise}.
    \end{cases}
\end{equation}
The coefficients on the added variables are time-varying, as before.

\begin{figure}[ht]
    \centering
    \subcaptionbox{East to West}
    {\includegraphics[width=0.45\textwidth]{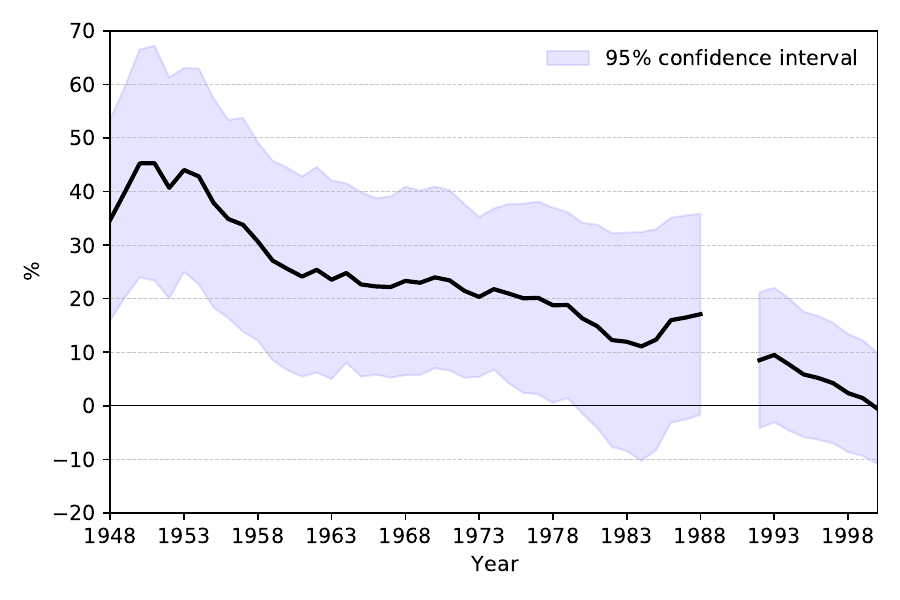}}
    \subcaptionbox{West to East}
    {\includegraphics[width=0.45\textwidth]{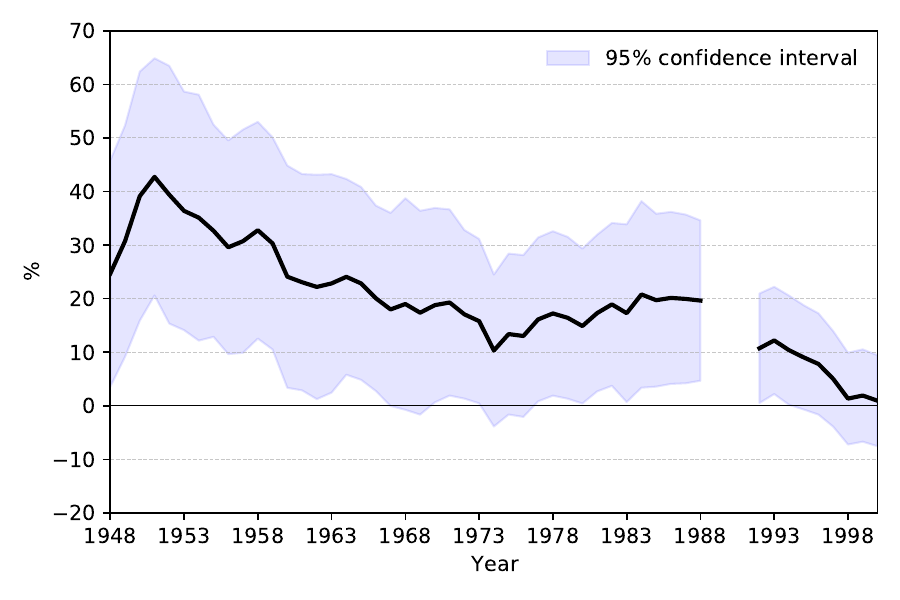}}

    \subcaptionbox{East to East}
    {\includegraphics[width=0.45\textwidth]{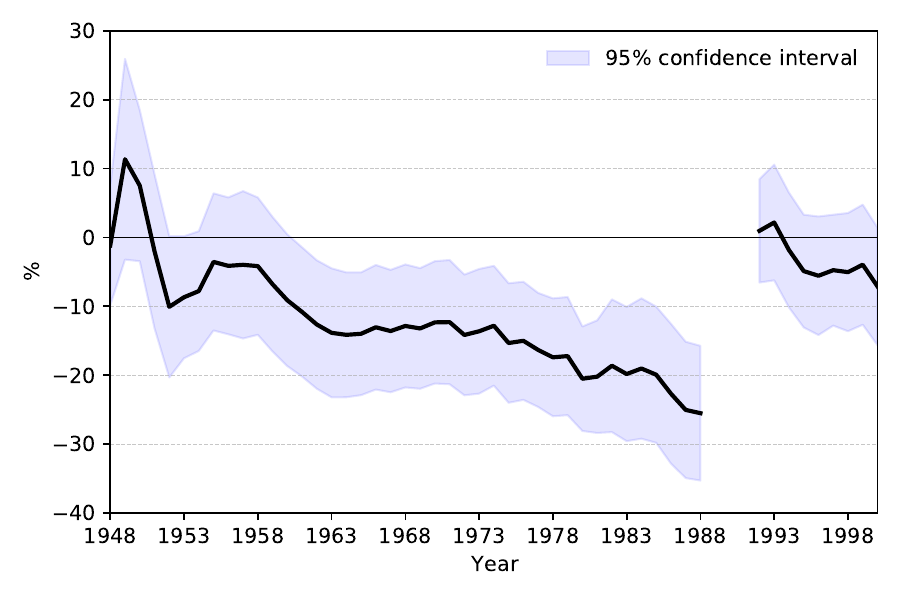}}
    \subcaptionbox{West to West}
    {\includegraphics[width=0.45\textwidth]{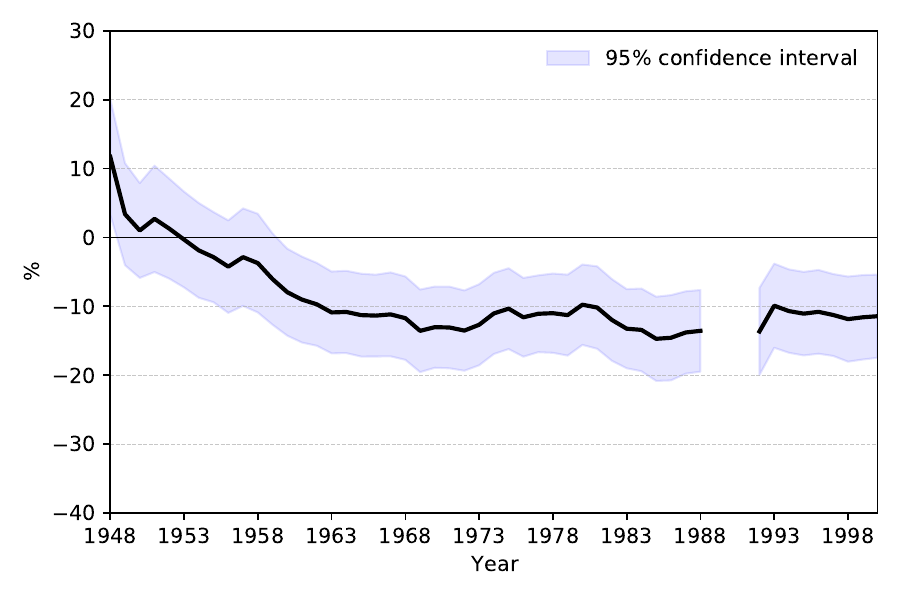}}
    \caption{Tariff equivalent of trade costs for flows across blocs and inside blocs} \label{fig:four_groups}
    \vspace{11pt}
    \begin{minipage}{0.9\textwidth}
    {\footnotesize \textbf{Notes}: The figures show the tariff equivalent of the trade costs estimated using the specification in~\eqref{eq:specification2}. The tariff equivalent measure is calculated from the estimates $\hat{\theta}_t^{ij}$, where $ij \in \{EW, WE, EE, WW\}$, using the transformation $100\times[\exp(-\hat{\theta}_t^{ij}/5.03)-1]$. The tariff equivalent measure is expressed in percentage points. The 95\% confidence interval is calculated using the delta method.}
    \end{minipage}
\end{figure}

The two figures showing trade flows across bloc borders (panels (a) and (b) in Figure~\ref{fig:four_groups}) are consistent with the Iron Curtain figure. Trade costs are higher and evolve gradually in the years before 1989. The point estimates of trade costs for exports from the West to the East suggest a relaxation of trade barriers in the mid-1980s, but the estimates for these years are very imprecise, as evidenced by the widening of the confidence interval for exactly these years. Looking at exports from West to East, there is a similar decline in trade costs in the early 1970s. In this case, the change in trade costs is not affected by less precision in the estimate, as the entire confidence interval for these years shifts.

Panels (c) and (d) in Figure~\ref{fig:four_groups} show three interesting phenomena. First, both graphs show a progressive decline in trade costs for intra-bloc trade until 1989. Second, the trade liberalization is three times greater for the Eastern bloc during this period. Third, after the fall of the Berlin Wall, the ease of intra-Eastern trade drops significantly and falls below that of intra-Western trade.

At first glance, the deep integration of the Eastern bloc countries might cast doubt on the uniqueness of the experience of economic integration resulting from the formation of the European Common Market and ultimately the European Union (EU), whose trade achievements are often cited as one of the most tangible benefits of the EU.

However, these findings are explained by the fact that while the Eastern bloc countries had very limited overall trade relations with the non-Soviet world, Western countries were generally more open and became increasingly so during the period when European institutions were being established. The simultaneous opening to trading partners outside Europe meant that the European integration process did not lead to significant trade diversion. This view is supported by a recent study by \citet{HeadMayer:21}, where the authors use a gravity model to analyze the evolution of trade costs among EU members, and between EU members and the rest of the world.

The sharp decline in the ease of trade between Eastern bloc countries after 1989 can probably be explained by the fall of the Berlin Wall, the dissolution of the Council for Mutual Economic Assistance, and ultimately the dissolution of the Soviet Union. The trade-integrating forces associated with these institutions were only gradually and partially replaced by other trade agreements, such as the Central European Free Trade Agreement, which had different objectives: not only to promote trade among the (now former) socialist states, but also to foster their integration with Western Europe.

%\clearpage

\textbf{Trade relations with non-aligned and neutral countries}. 
We now extend the analysis to examine each bloc's trade relations with neutral countries in Europe. We distinguish between three groups of countries: countries that are officially neutral but lean toward the West (Austria, Ireland, and Sweden), countries that are actually neutral (Switzerland and Finland), and Yugoslavia, a famously non-aligned country.

We proceed as before and add dummy variables for trade between these three groups of countries and the Eastern and Western blocs. We also add dummy variables for intra-bloc trade whenever appropriate (i.e., in all cases except Yugoslavia, which is a single country).

\begin{figure}[htbp]
    \centering
    \subcaptionbox{Trade with the East}
    {\includegraphics[width=0.45\textwidth]{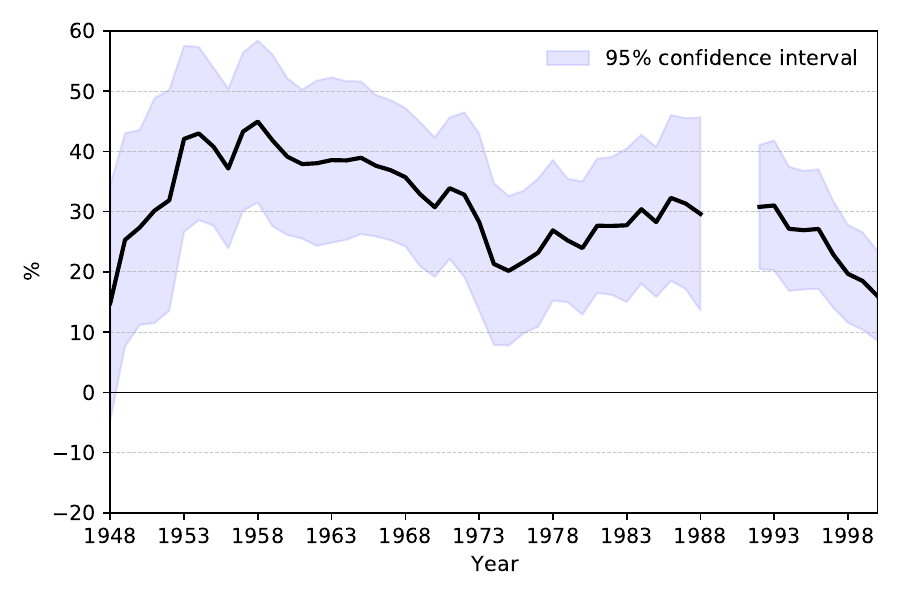}}
    \subcaptionbox{Trade with the West}
    {\includegraphics[width=0.45\textwidth]{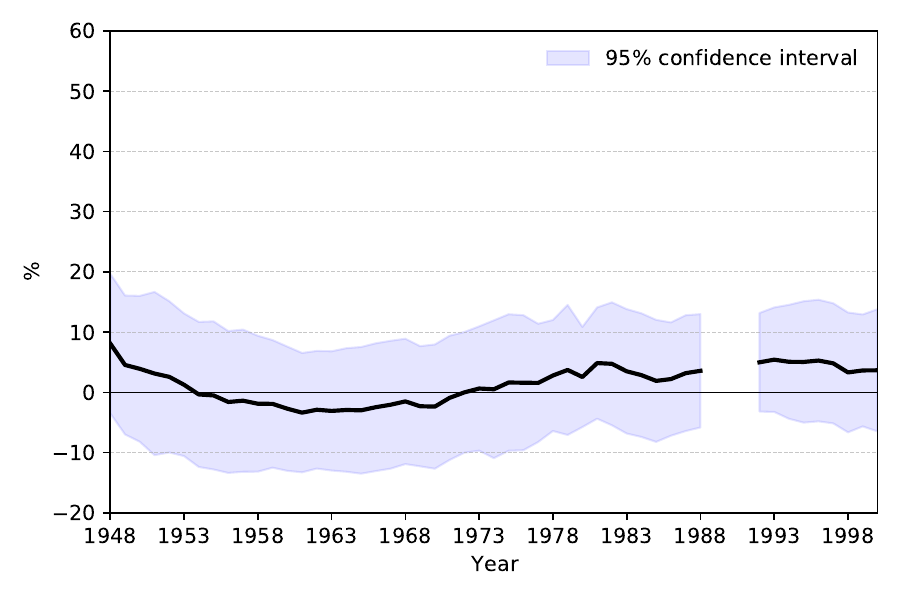}}
    \caption{Tariff equivalent of the border for West-leaning countries} \label{fig:L}
    \vspace{11pt}
    \begin{minipage}{0.9\textwidth}
    \footnotesize
    \textbf{Notes}: The figures show the tariff equivalent of the trade costs for West-leaning countries (Switzerland, Ireland, and Sweden). The tariff equivalent measure is expressed in percentage points. The 95\% confidence interval is calculated using the delta method.
    \end{minipage}
\end{figure}

\begin{figure}[ht]
    \centering
    \subcaptionbox{Trade with the East}
    {\includegraphics[width=0.45\textwidth]{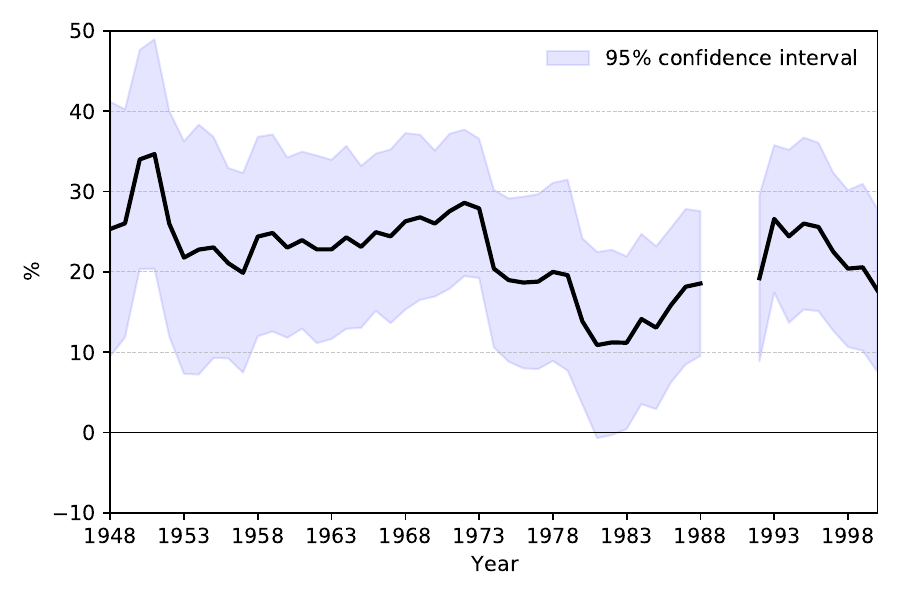}}
    \subcaptionbox{Trade with the West}
    {\includegraphics[width=0.45\textwidth]{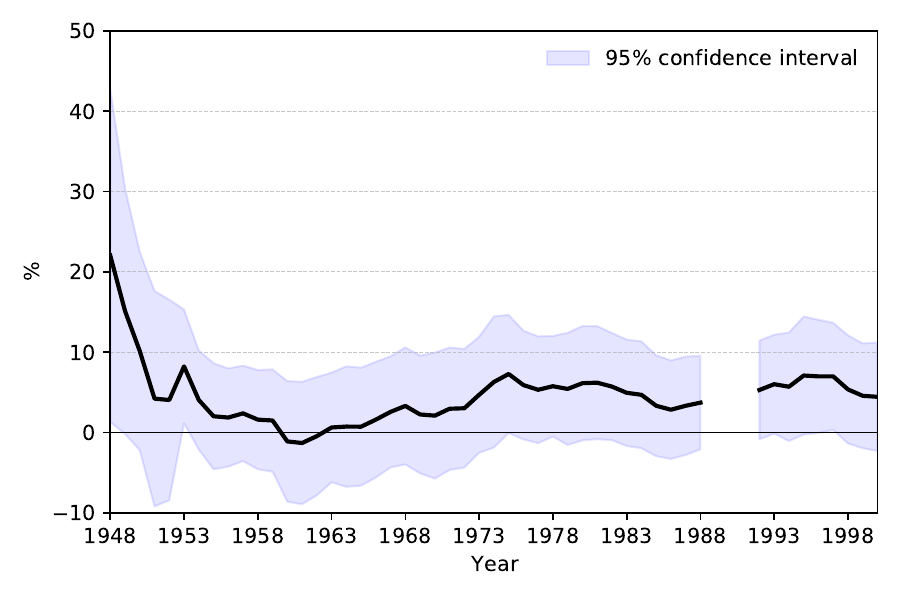}}
    \caption{Tariff equivalent of the border for neutral countries} \label{fig:N}
    \vspace{11pt}
    \begin{minipage}{0.9\textwidth}
    \footnotesize
    \textbf{Notes}: The figures show the tariff equivalent of the trade costs for neutral countries (Austria and Finland). The tariff equivalent measure is expressed in percentage points. The 95\% confidence interval is calculated using the delta method.
    \end{minipage}
\end{figure}

\begin{figure}[ht]
    \centering
    \subcaptionbox{Trade with the East}
    {\includegraphics[width=0.45\textwidth]{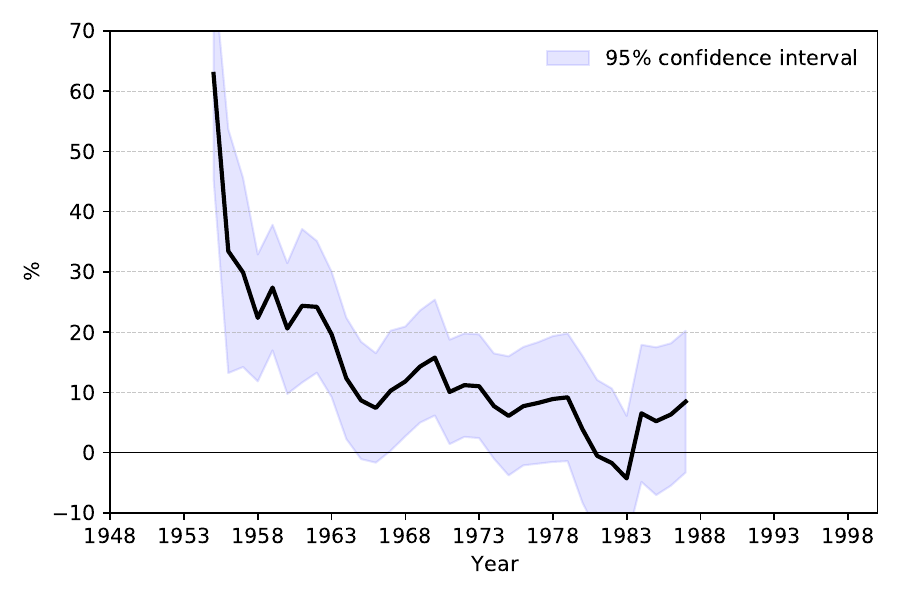}}
    \subcaptionbox{Trade with the West}
    {\includegraphics[width=0.45\textwidth]{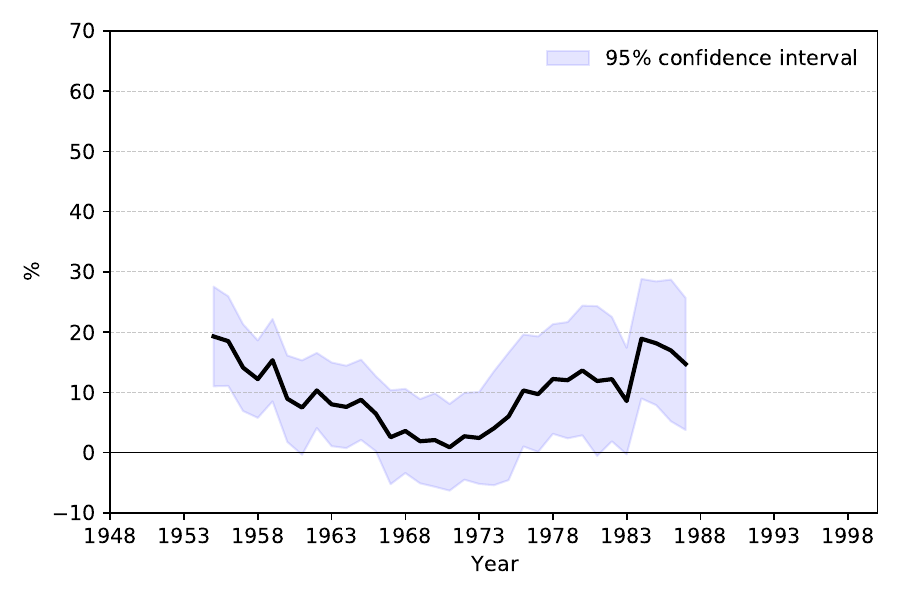}}
    \caption{Tariff equivalent of the border for Yugoslavia} \label{fig:Y}
    \vspace{11pt}
    \begin{minipage}{0.9\textwidth}
    \footnotesize
    \textbf{Notes}: The figures show the tariff equivalent of the trade costs for Yugoslavia. Estimates are not available for all years because of lack of data. The tariff equivalent measure is expressed in percentage points. The 95\% confidence interval is calculated using the delta method.
    \end{minipage}
\end{figure}

The Western-leaning countries experienced increased barriers to trade with the East at the same level as with the West, while trade with the West was not hindered by increased barriers. Austria and Finland had very similar levels of trade barriers with the East as the West-leaning countries, at least until the early 1970s. This suggests that the distinction between westward-leaning and neutral countries is not substantial in terms of trade relations.

Interestingly, both Western-leaning and neutral countries show similar patterns in the evolution of calculated ad valorem tariff equivalents. In particular, after being very high at the beginning of the Cold War period, ad valorem tariff equivalents begin to decline around the late 1960s and until the late 1970s, corresponding to the ``détente''--a period of reduced tensions between the United States and the Soviet Union \citep{VillaumeWestad:10}.

During this period, the enforcement of the Western export control system was relaxed, creating fewer difficulties for exporters in Western-leaning and neutral countries. They could now more easily re-export Western goods to the East. In the early 1980s, ad valorem tariff equivalents began to rise again, coinciding with renewed U.S. pressure on these countries to tighten controls on re-exports \citep{Jensen-Eriksen:19}.

Yugoslavia, on the other hand, experienced increased trade costs with both blocs. Our estimates capture the almost complete break in trade relations between Yugoslavia and the Eastern bloc after the Soviet-Yugoslav split---Yugoslavia's departure from the Soviet sphere of influence in 1948\footnote{In 1961, Yugoslavia became a founding member of the Non-Aligned Movement, a group of countries that were not formally aligned neither with nor against either the Western or Eastern bloc.}---and its reversal in the 1950s \citep{McKenzie:08}.

Together with the previous results, this implies that the Eastern bloc countries were the most disconnected from the rest of the world, as they experienced increased trade costs not only with the West, but also with non-aligned or neutral countries.

%\clearpage

\section{A quantification of missing trade and welfare losses due to the Iron Curtain} \label{sec:quantification}

In this section we quantify the impact of the Iron Curtain on trade and welfare by considering a counterfactual world without the Iron Curtain. For our counterfactual simulations, we use a standard  general equilibrium trade model with a production function that leads to a positive supply elasticity. As is well known \citep[cf.][]{HeadMayer:14}, the computation of general equilibrium counterfactuals requires only knowledge of bilateral trade flows and a few sufficient statistics. In this case, we need two elasticities: the trade elasticity, which measures how bilateral trade flows respond to a change in bilateral trade costs, all else equal, and the supply elasticity, which measures how output in a country responds to an increase in the relative price of its export good. 

\textbf{Brief description of the model}. The quantitative model %we use for the simulations 
is a standard trade model in which goods are produced by combining labor with intermediate
inputs. This feature of the production side of the model, called ``roundabout production'', leads to a positive supply elasticity, as in the general class of trade models analyzed by \citet{AllenArkolakisTakahashi:20}. Trade is costly and is characterized by ad valorem iceberg trade costs. Demand is given by CES preferences defined over varieties differentiated by origin, i.e., it follows the models of \citet{Armington:69}, \citet{Anderson:79} and \citet{AndersonvanWincoop:03}. As demonstrated by \cite{ArkolakisCostinotRodriguez-Claire:12} and \cite{AllenArkolakisTakahashi:20}, for a given set of parameters, this model is isomorphic in terms of its trade and welfare implications to a wide class of alternative trade models. We give a more detailed description in the model in Appendix~\ref{sec:appendix_model}.

\textbf{Choice of elasticities}. We take the trade elasticity from the value reported in the handbook chapter by \citet{HeadMayer:14} and set it to 5.03. This is also the trade elasticity we used to express the effect of the Iron Curtain in tariff equivalent terms. The supply elasticity depends on the importance of intermediates in the production function. We follow the strategy of \citet{CamposEstefania-FloresFurceriTimini:23} and choose the supply elasticity as the midpoint of the supply elasticities implied by the 10th and 90th percentiles of the distribution of the range of intermediate shares for the sample of countries in the KLEMS database, as reported by \citet{HuoLevchenkoPandalai:23}. This yields a supply elasticity of 1.24, which is slightly higher than the value of 1.0 used by \citet{AlvarezLucas:07}, but lower than 3.76, the value favored by \citet{EatonKortum:02} in their work. Since the model is static, we solve the model for a counterfactual equilibrium in each year starting in 1950.\footnote{East Germany is missing domestic trade for the years before 1954. For the simulations, we extrapolated domestic trade by regressing log(GDP) on a linear trend for East Germany and imputing the data with fitted values. For this country, we also completed the bilateral data after 1974 with imputed trade flows. Appendix~A explains how data are imputed for this country.}

\textbf{Simulation exercise}. The simulation exercise consists of simulating a counterfactual scenario in which the Iron Curtain does not exist, or at least does not raise trade barriers between East and West. In each year, we remove the increased trade costs between East and West and solve for counterfactual trade flows and welfare using exact hat algebra methods. The algorithm we employ is described by \citet{CamposReggioTimini:24}.

\begin{figure}[ht]
    \centering
    \subcaptionbox{Volume}
    {\includegraphics[width=0.45\textwidth]{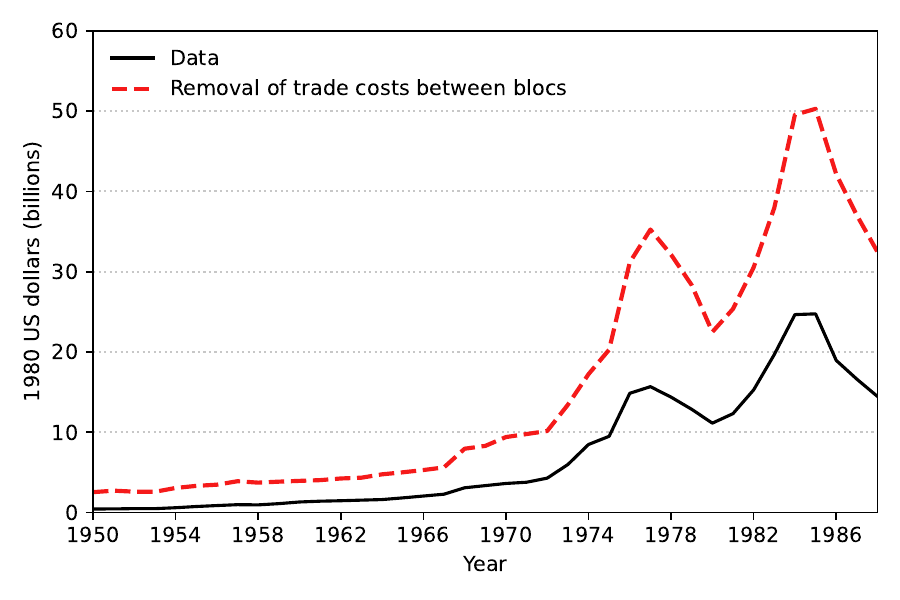}}
    \subcaptionbox{Percent differences}
    {\includegraphics[width=0.45\textwidth]{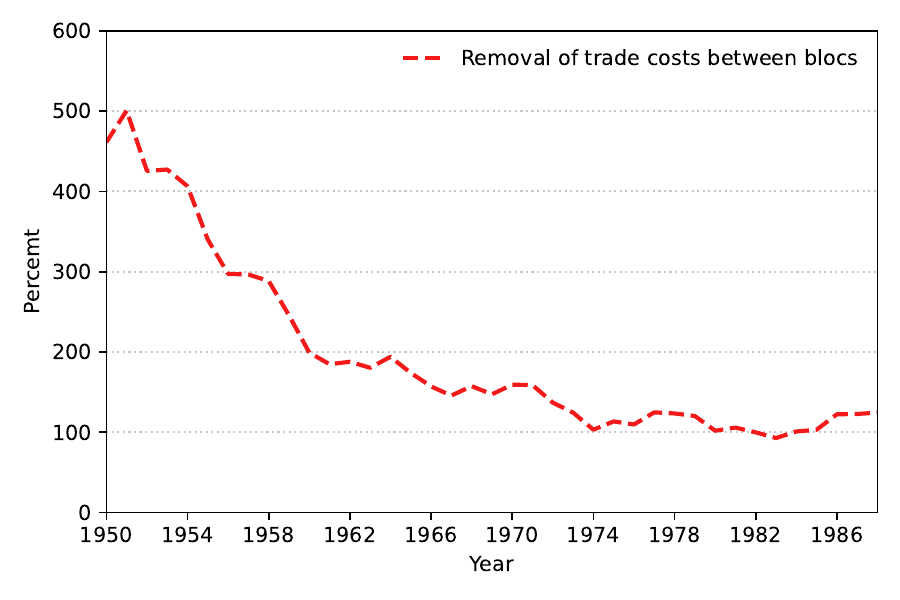}}
    \caption{Trade flows between East and West} \label{fig:counterfactual_EW}
    \vspace{11pt}
    \begin{minipage}{0.9\textwidth}
    \footnotesize
    \textbf{Notes}: The figure shows the results of a simulation in which the trade barriers of the Iron Curtain are removed. The solid line in the left panel shows the actual trade volume between East and West. The dashed line shows the counterfactual trade volume. The panel on the right shows the predicted percentage increase in trade volume that would occur if the trade barriers of the Iron Curtain were removed.
    \end{minipage}
\end{figure}

\textbf{Results}. In Figure~\ref{fig:counterfactual_EW} we show the counterfactual level of trade flows between East and West. The left panel shows the trade flows (sum of both directions) and the right panel shows the percentage change from what was observed in the data.

Through the lens of our quantitative model, the presence of the Iron Curtain roughly halved East-West trade flows. In relative terms, the trade lost to the Iron Curtain was greatest at the beginning of the Cold War, when inter-bloc trade would have been five times higher if the effect of the Iron Curtain had been removed. This relative measure of the importance of the Iron Curtain declined over time. By the end of the Cold War, the trade lost to the Iron Curtain was about 100\%, i.e., the model predicts that inter-bloc trade would double if the Iron Curtain were removed.

\begin{figure}[ht]
    \centering
    \subcaptionbox{Eastern countries}
    {\includegraphics[width=0.45\textwidth]{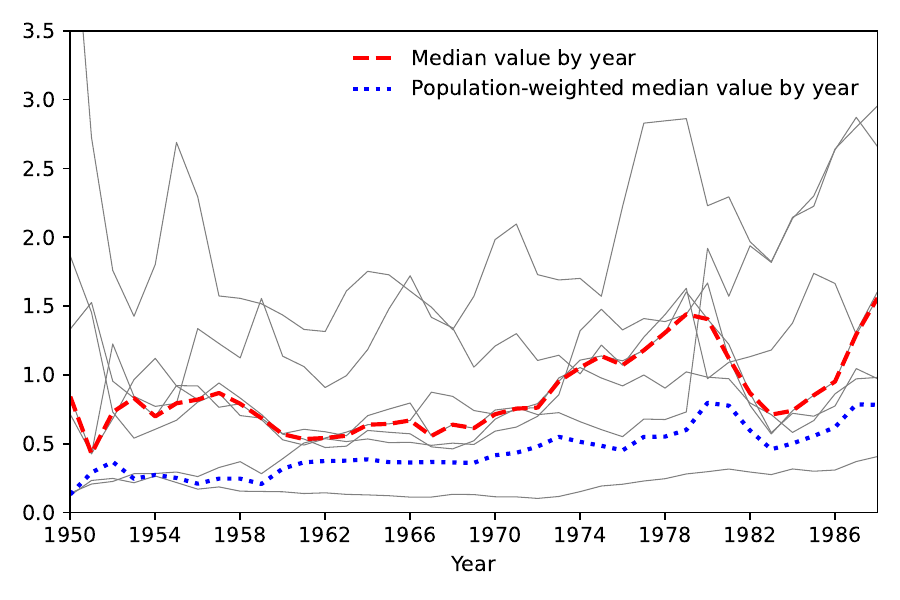}}
    \subcaptionbox{Western countries}
    {\includegraphics[width=0.45\textwidth]{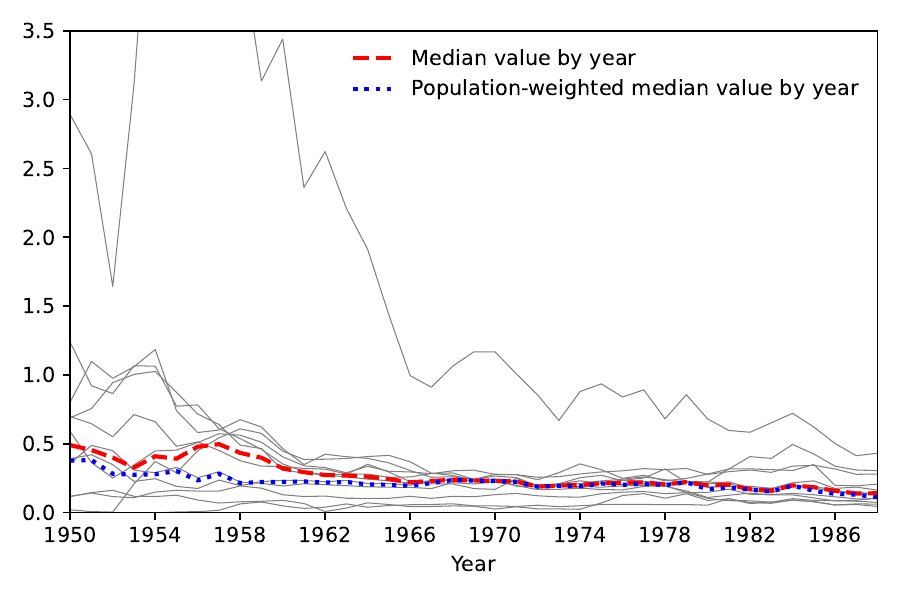}}
    \caption{Welfare gains from removing trade barriers between blocs} \label{fig:counterfactual_EW_welfare}
    \vspace{11pt}
    \begin{minipage}{0.9\textwidth}
    \footnotesize
    \textbf{Notes}: The figure shows the results from a simulation in which the trade barriers due to the Iron Curtain are removed. The individual thin lines welfare gains from removing trade costs between blocs for all countries in each bloc. The dashed line shows the median welfare gains by year. The dotted line shows the population-weighted mean.
    \end{minipage}
\end{figure}

The impact of the removal of trade barriers on welfare is shown in Figure~\ref{fig:counterfactual_EW_welfare}, where we plot the change in welfare for each country in each bloc over time. According to the theoretical model used to derive these results, the change in welfare can be interpreted as the change in the per capita consumption of a representative agent or the change in the real wage in each country expressed in terms of the consumption bundle.

Eastern countries tend to be more affected by the change in trade costs, as expected, since they are generally more closed to the rest of the world, except for other countries in their same bloc. The median country in the Eastern bloc would have experienced a welfare gain of nearly 1\% per year if the trade costs of the Iron Curtain had been eliminated.

Western countries, on the other hand, would have experienced welfare gains well below the 1\% mark. The only exception is Iceland, which---as a small country---is a clear outlier in the group of Western countries shown in panel~(b) of ~\ref{fig:counterfactual_EW_welfare}. The welfare gain from eliminating the trade barriers associated with the Iron Curtain is clearly decreasing over time, reflecting both the increase in the size of the Western European economy and the decrease in the tariff equivalent measure of the Iron Curtain. This implies that the Iron Curtain became less of a burden for Western Europe in the later years of the Cold War.

This finding for Western countries contrasts with the important role the Iron Curtain continued to play for countries in the East. Not only did the welfare losses associated with the existence of the Iron Curtain persist until the end of the Cold War, but they also tended to increase for the median country. This suggests that welfare losses due to the lack of international trade may have been behind the drive to liberalize the East bloc economies at the end of the Cold War, such as the implementation of perestroika in the Soviet Union, which allowed ministries to act more independently and introduced market-like reforms in an attempt to adopt some of the characteristics of Western economies.

\clearpage

\section{Conclusion} \label{sec:conclusion}
The Iron Curtain---a symbolical and physical barrier dividing Europe
into two distinct areas---was an important driver of trade barriers that had large effects on trade between East and West and the welfare of nations during the Cold War.

In this paper, we quantify the evolution of a tariff-equivalent measure of the Iron Curtain using previously unavailable data that include bilateral trade flows for East Germany and the USSR. We also analyze how the Iron Curtain generated trade integration within the East and West blocs and how trade barriers with non-aligned or neutral countries evolved over time.

Using a quantitative trade model, we show that despite the gradual easing of trade restrictions over time, the Iron Curtain still had a significant impact on trade flows and welfare, especially in the Eastern bloc. The division created by the Iron Curtain also led to an increase in intra-bloc trade, especially in the Eastern bloc. In terms of per capita real wages, or welfare, the Iron Curtain led to persistent losses in the Eastern bloc countries of about 1\% per year until the end of the Cold War, despite a decline in the tariff-equivalent measure of these barriers. These losses suggest a possible driving force behind efforts to liberalize the Eastern bloc economies, including the implementation of perestroika and other market-oriented changes in the Soviet Union.

In conclusion, the Iron Curtain served as a formidable barrier to trade between Eastern and Western countries, illustrating the perils of geopolitical fragmentation.

\clearpage
\end{doublespace}

\bibliographystyle{ecta}
\bibliography{iron-curtain}

\clearpage
\appendix
\begin{center}
{\sc \Large Appendix}
\end{center}

\counterwithin{figure}{section}
\counterwithin{table}{section}

\section{Historical data on bilateral trade for East Germany and the USSR}
The bilateral trade data come from the freely available TRADHIST database (version 4). The original provenance for more than 95\% of the data in the period we consider (1948--2000) is the Direction of Trade Statistics (DOTS) database produced by the International Monetary Fund. We augment the data set with bilateral trade data for East Germany and USSR. In doing so, we perform the same processing steps used by the IMF for the DOTS database, as set out in their methodology.

\subsection{General Methodology}
\textbf{Data processing}. Using the methodological guide for DOTS \citep{IMF:1993}, we replicate the exact steps that the IMF would have performed if it had had access to the data in the statistical yearbooks.
Therefore, to convert the values from local currency into US dollars, we use the exchange rates as reported in the International Financial Statistics (IFS), or the official exchange rate if it is unavailable. This complies with Section 3.1.1 in the methodology. We convert trade flows in local currencies to US dollars (the unit of account of DOTS). We also make the recommended adjustment for c.i.f./f.o.b discrepancies. In the DOTS database, exports should be recorded using f.o.b. values and imports using c.i.f. values. Section 3.1.2 part~(2) recommends using a factor of 10\% to increase f.o.b. import values to obtain the c.i.f. value. We verify that export values obtained in this way are virtually identical to the values reported in DOTS for the years in which they are available for East Germany and the Soviet Union. We have access to several years of statistical yearbooks/reports. In accordance with the general practice of DOTS, we update estimates to the latest published value when there are revisions from one year to the next.

\textbf{Incorporation of new data in the TRADHIST database}. Values in the TRADHIST database are expressed in British pounds. We use the British pound/US-dollar exchange rate in the IFS to convert dollar values into British pounds.

The TRADHIST database always prefers data from primary sources to data from secondary sources. It also prefers the value of a bilateral trade flow as reported by the importing country. In consequence, we replace observations in the TRADHIST database with the data we collected for all trade flows that are an import of East Germany or the USSR. We replace data using East German or USSR exports only when an observation in TRADHIST is either missing or zero and our collected data shows positive flows.

In the resulting completed database we clearly signal which observations have been updated by overwriting the values in the variable \texttt{SOURCE\_TF}. We use the code \texttt{"EDEU\_IP"} if the original source is an import flow reported in the statistical yearbook of East Germany (1,103 cases), \texttt{"EDEU\_XP"} if it is an export flow in the statistical yearbook of East Germany (403 cases),
\texttt{"USSR\_IP"} if it is an import flow in the statistical report of the USSR (2,914 cases), and \texttt{"USSR\_XP"} if it is an export flow in the statistical report of the USSR (676 cases).

\subsection{Sources and details for East Germany}
We obtain bilateral trade flows reported in the statistical yearbook (Statistisches Jahrbuch der Deutschen Demokratischen Republik in the original), which has been digitized and is available at DigiZeitschriften, a German online journal archive. The current link to the statistical yearbooks is \url{https://www.digizeitschriften.de/search?q=514402644} An archived snapshot of this web page for 6 October 2023 has been preserved at: \url{https://web.archive.org/web/20231006162901/https://www.digizeitschriften.de/search?q=514402644}.

Trade data in the the statistical yearbooks are reported in Valuta-Mark, a unit of account used by East Germany for international trade that is nominally equivalent to the Western Deutsche Mark. Both exports and imports are reported using a free on board (f.o.b.) valuation.\footnote{The statement in original language that indicates this in the methodology is the following: ``Die Werte enthalten den Warenpreis zuzüglich aller Fracht- und Nebenkosten im Lieferland (frei Grenze Lieferland bzw. fob Verschiffungshafen).''} We first convert the data into US dollars and then to British pounds using the exchange rates in the IFS.

The statistical yearbooks only record bilateral imports and exports separately until 1974. After that year they report only total trade (i.e., the sum of exports and imports) for each country pair. For these later years, if a trade flow is missing in TRADHIST, but the trade flow in the opposite directions is available in TRADHIST, then we obtain the missing observation by subtracting from total trade the value of the observation in the opposite direction.
In the database, we identify these cases by setting the variable \texttt{SOURCE\_TF} to \texttt{"EDEU\_TOTAL"} (63 cases).

When no observation in either direction is available, then we construct a measure of imputed trade flows in both directions by splitting the total trade value in half. In the database, we create a new variable (\texttt{FLOW\_IMPUTED}) with this measure and do not overwrite the data. We calculate these imputed values in case they are useful for other researchers, but we do not use them in the estimations in this paper.

\subsection{Sources and details for the USSR}
We obtain bilateral trade flows reported in the yearly foreign trade statistical reviews for the USSR (Внешняя торговля СССР (Статистический обзор) in the original), which have been digitized and are available at \url{https://istmat.org/taxonomy/term/344}. An archived snapshot of this web page for 24 June 2023 has been preserved at: \url{https://web.archive.org/web/20230624083918/https://istmat.org/taxonomy/term/344}.

Trade data in the statistical reports from the Soviet Union are reported in rubles. Both exports and imports are reported using a free on board (f.o.b.) valuation.\footnote{The statement in original language in the methodology that indicates this is the following: Стоимость товаров подсчитывается по ценам контрактов, приведенным к единому базису, а именно: по экспорту - к ценам фоб советские порты или франко-сухопутная граница СССР; по импорту - к ценам фоб иностранные порты или франко-граница страны отгрузки.
Пересчет иностранных валют в рубли произведен по официальному курсу Государственного банка СССР за соответствующий период.} We convert values in rubles to US dollars using the official exchange rate between the ruble and the US dollar from an archived spreadsheet that was originally available at the web page of the central bank of the Russian Federation.\footnote{Trade values are converted into US dollars DOTS using official rates when market values are not available in the IFS database.} The permanent link to an archived version of this spreadsheet (3 February 2013) is the following: \url{https://web.archive.org/web/20130203063808/http://www.cbr.ru/currency_base/OldDataFiles/USD.xls}. To convert values from US dollars to British pounds we use the exchange rate in the IFS.

\subsection{Validation}
We compare our hand-collected data with the data in TRADHIST that comes from the DOTS database and in which the reporting country is the importer. We show figures for imports of East Germany from West Germany, the Soviet Union, and the USA, and for imports of the Soviet Union from West Germany, the Soviet Union, and the USA. As can be seen in the figures, observations are virtually identical when they overlap.

\begin{figure}[ht]
\begin{center}
\includegraphics[width=0.7\textwidth]{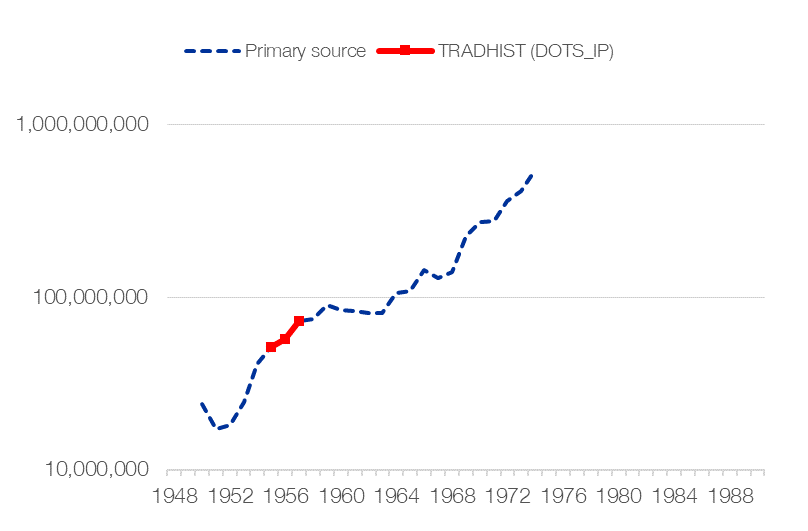}
\vspace{11pt}
\begin{minipage}{0.9\textwidth}
\caption{Trade flows from West Germany to East Germany}
\label{fig:validation_1}
\footnotesize \textbf{Notes}: Values are expressed in pounds sterling. The vertical axis uses a logarithmic (base 10) scale. Values from the TRADHIST database derived from IMF DOTS are plotted with a red line with squares. Data collected and processed by us are plotted with a blue dashed line.
\end{minipage}
\end{center}
\end{figure}

\begin{figure}[ht]
\begin{center}
\includegraphics[width=0.7\textwidth]{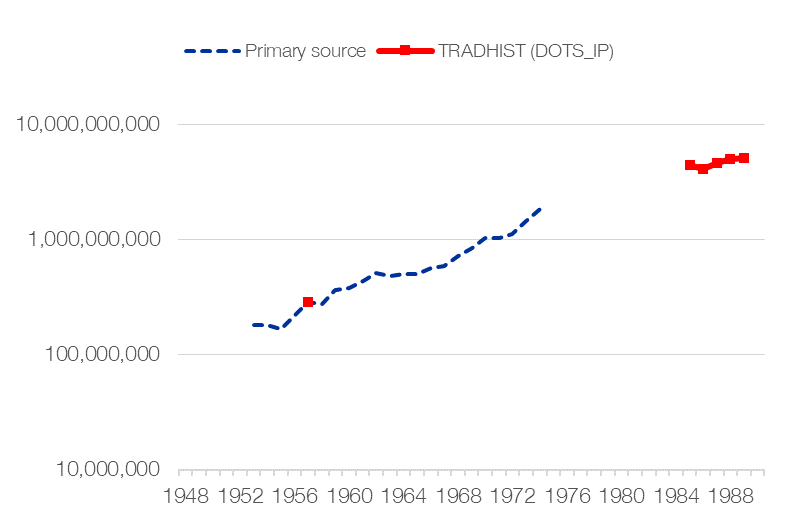}
\vspace{11pt}
\begin{minipage}{0.9\textwidth}
\caption{Trade flows from the USSR to East Germany}
\label{fig:validation_2}
\footnotesize \textbf{Notes}: Values are expressed in pounds sterling. The vertical axis uses a logarithmic (base 10) scale. Values from the TRADHIST database derived from IMF DOTS are plotted with a red line with squares. Data collected and processed by us are plotted with a blue dashed line.
\end{minipage}
\end{center}
\end{figure}

\begin{figure}[ht]
\begin{center}
\includegraphics[width=0.7\textwidth]{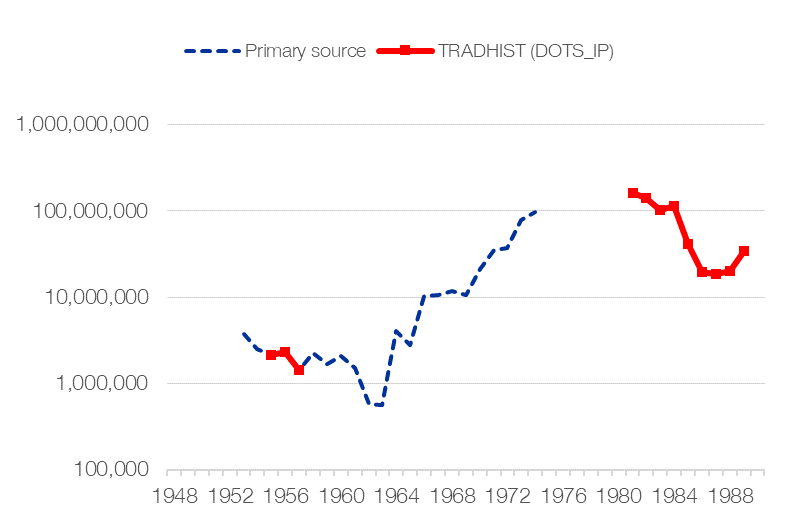}
\vspace{11pt}
\begin{minipage}{0.9\textwidth}
\caption{Trade flows from the USA to East Germany}
\label{fig:validation_3}
\footnotesize \textbf{Notes}: Values are expressed in pounds sterling. The vertical axis uses a logarithmic (base 10) scale. Values from the TRADHIST database derived from IMF DOTS are plotted with a red line with squares. Data collected and processed by us are plotted with a blue dashed line.
\end{minipage}
\end{center}
\end{figure}

\begin{figure}[ht]
\begin{center}
\includegraphics[width=0.7\textwidth]{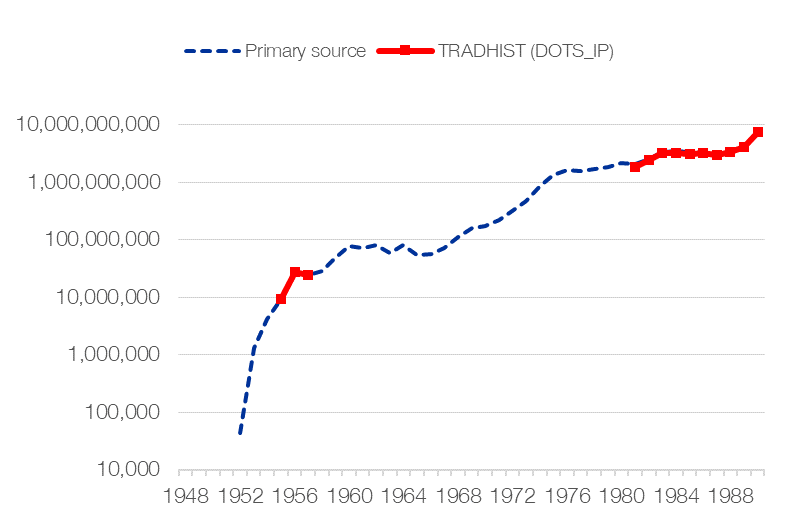}
\vspace{11pt}
\begin{minipage}{0.9\textwidth}
\caption{Trade flows from West Germany to the USSR}
\label{fig:validation_4}
\footnotesize \textbf{Notes}: Values are expressed in pounds sterling. The vertical axis uses a logarithmic (base 10) scale. Values from the TRADHIST database derived from IMF DOTS are plotted with a red line with squares. Data collected and processed by us are plotted with a blue dashed line.
\end{minipage}
\end{center}
\end{figure}

\begin{figure}[ht]
\begin{center}
\includegraphics[width=0.7\textwidth]{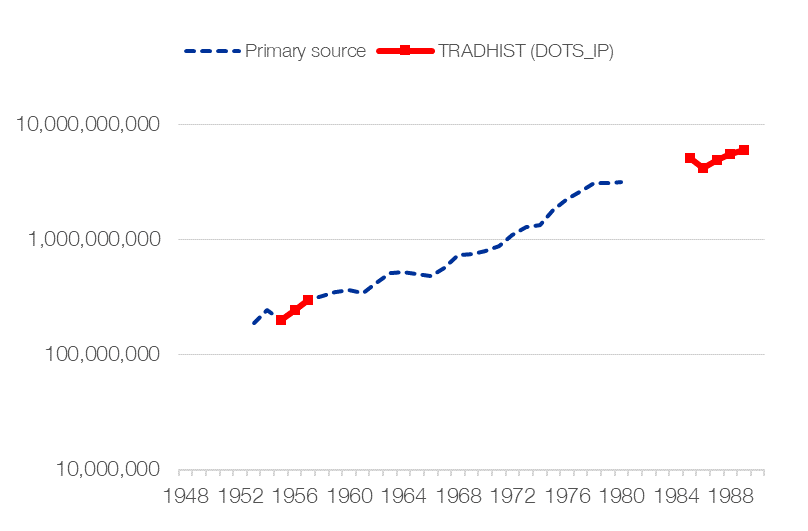}
\caption{Trade flows from East Germany to the USSR}
\label{fig:validation_5}
\vspace{11pt}
\begin{minipage}{0.9\textwidth}
\footnotesize \textbf{Notes}: Values are expressed in pounds sterling. The vertical axis uses a logarithmic (base 10) scale. Values from the TRADHIST database derived from IMF DOTS are plotted with a red line with squares. Data collected and processed by us are plotted with a blue dashed line.
\end{minipage}
\end{center}
\end{figure}

\begin{figure}[ht]
\begin{center}
\includegraphics[width=0.7\textwidth]{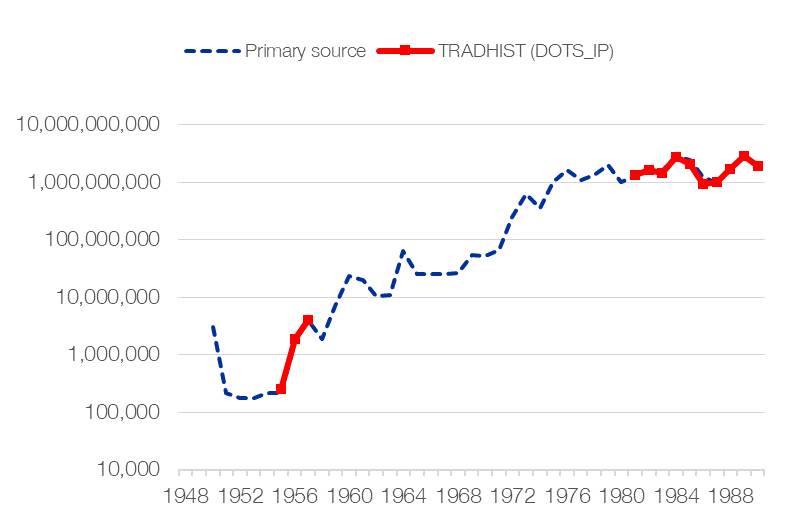}
\caption{Trade flows from the USA to the USSR}
\label{fig:validation_6}
\vspace{11pt}
\begin{minipage}{0.9\textwidth}
\footnotesize \textbf{Note}:Values are expressed in pounds sterling. The vertical axis uses a logarithmic (base 10) scale. Values from the TRADHIST database derived from IMF DOTS are plotted with a red line with squares. Data collected and processed by us are plotted with a blue dashed line.
\end{minipage}
\end{center}
\end{figure}

\clearpage
\section{Empirical appendix} \label{sec:appendix_empirical}

\begin{figure}[ht]
\begin{center}
    \includegraphics[width=0.9\textwidth]{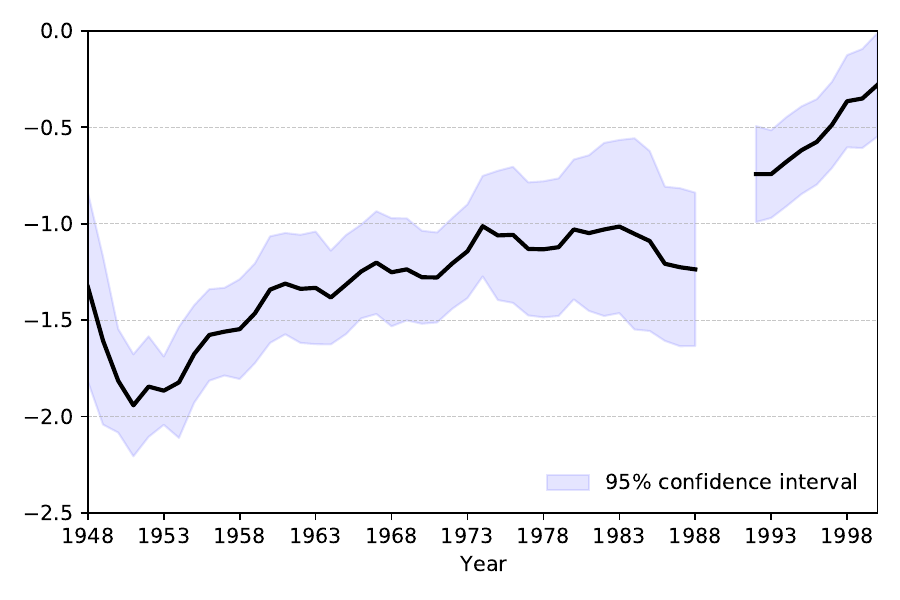}
\caption{Estimated coefficients of the Iron Curtain}
\label{fig:raw_estimates}
\vspace{11pt}
\begin{minipage}{0.9\textwidth}
\footnotesize
\textbf{Notes}: The figure shows the estimated coefficient of the Iron Curtain's borders ($\hat{\theta}_t$). The estimation uses the specification in~\eqref{eq:specification}. Standard errors are clustered by exporter, importer, and year. 
\end{minipage}
\end{center}
\end{figure}

Robustness check that estimates the baseline specification but adds intra-bloc trade.

\begin{figure}[ht]
    \centering
    \subcaptionbox{East to West}
    {\includegraphics[width=0.45\textwidth]{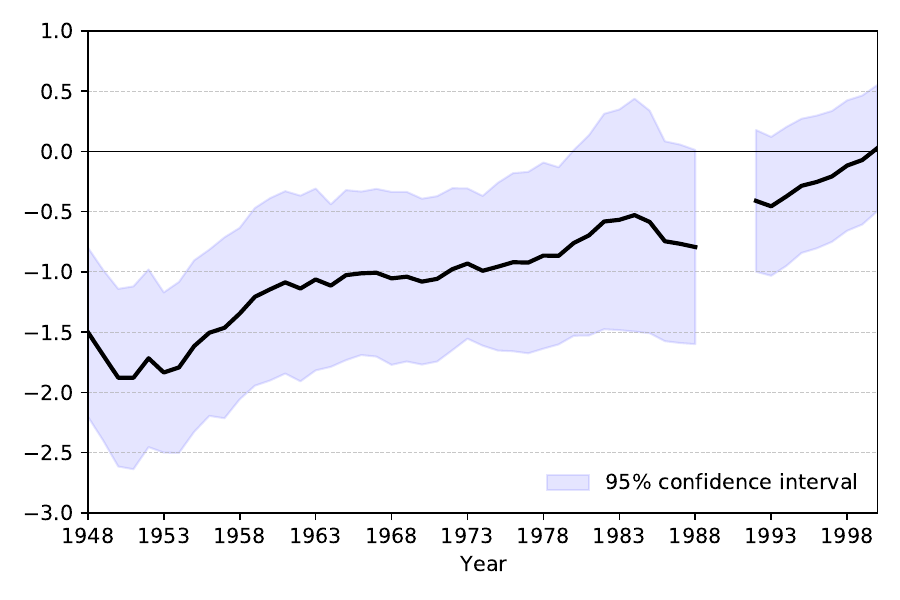}}
    \subcaptionbox{West to East}
    {\includegraphics[width=0.45\textwidth]{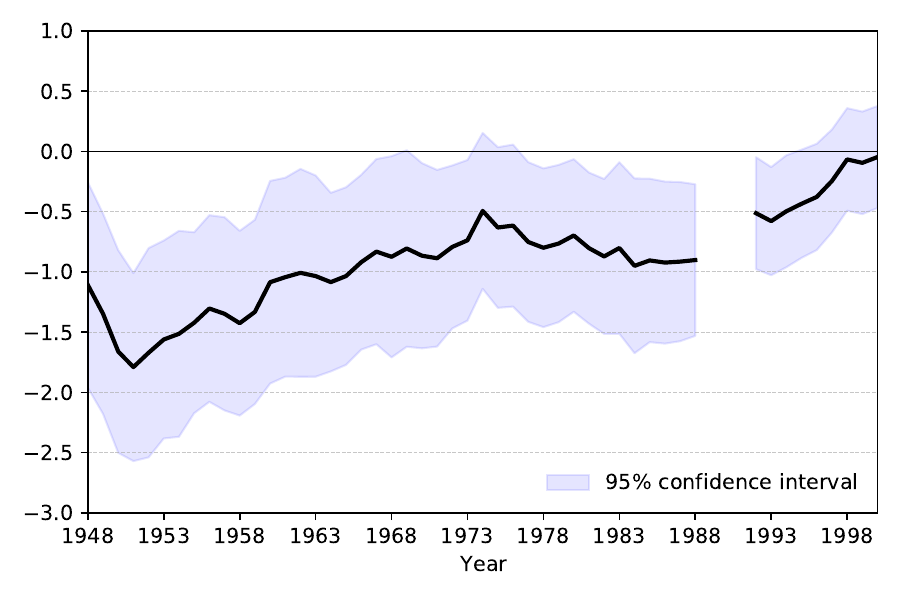}}

    \subcaptionbox{East to East}
    {\includegraphics[width=0.45\textwidth]{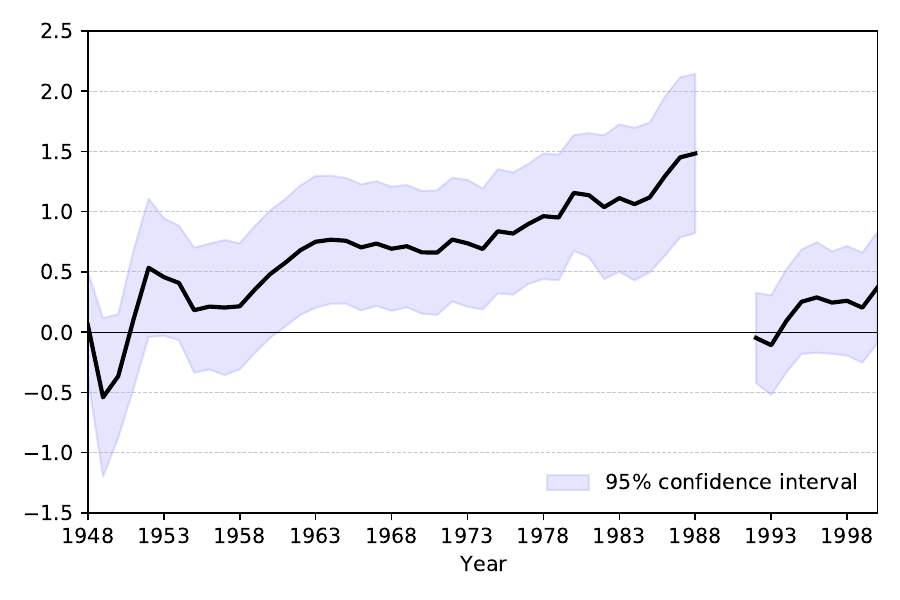}}
    \subcaptionbox{West to West}
    {\includegraphics[width=0.45\textwidth]{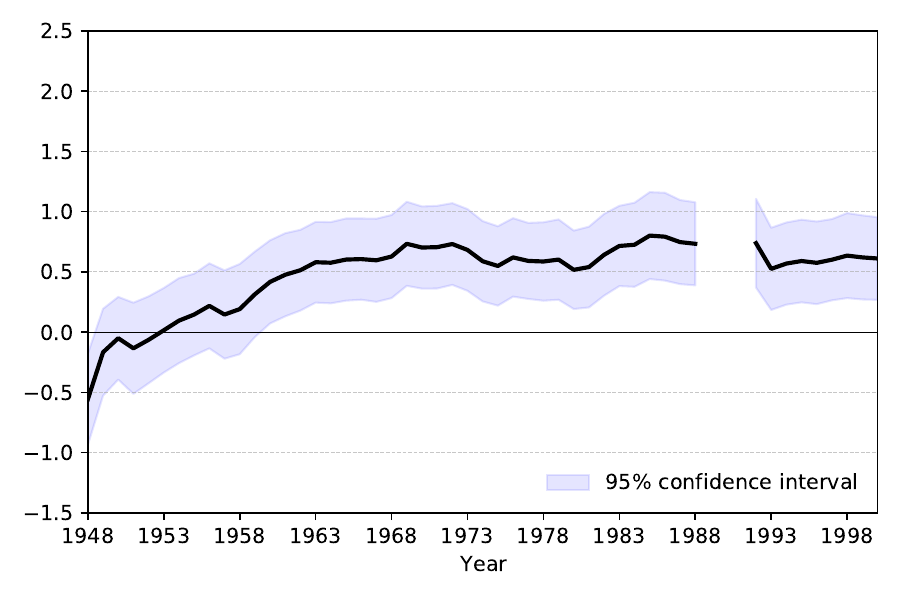}}
    \caption{Trade costs for flows across blocs and inside blocs} 
    \vspace{11pt}
    
    \begin{minipage}{0.9\textwidth}
    \footnotesize
    \textbf{Notes}: The figures show the estimated coefficients from the specification in~\eqref{eq:specification2}. Standard errors are clustered by exporter, importer, and year. 
    \end{minipage}
\end{figure}

\begin{figure}[ht]
\begin{center}
    \includegraphics[width=0.9\textwidth]{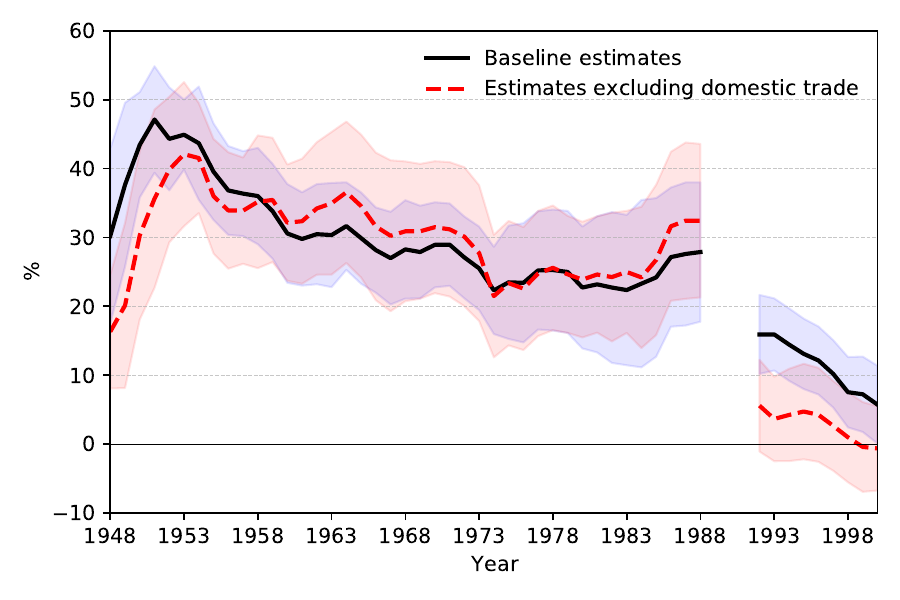}
\caption{Estimated tariff equivalent of the Iron Curtain with and without domestic trade}
\label{fig:robustness_domestic}
\vspace{11pt}
\begin{minipage}{0.9\textwidth}
{\footnotesize 
\textbf{Note}: The figure shows the estimated tariff equivalent of the Iron Curtain borders in percentage points. The estimation uses the specification in~\eqref{eq:specification}. The solid line shows results from the baseline estimation, which includes domestic trade. The dashed line shows results from an exercise in which all observations involving domestic trade are dropped from the estimation. The tariff equivalent is calculated from the estimates of $\hat{\theta}_t$ using the transformation $100\times[\exp(-\hat{\theta}_t/5.03)-1]$. The 95\% confidence interval is calculated using the delta method.}
\end{minipage}
\end{center}
\end{figure}

\clearpage

\section{Quantitative trade model} \label{sec:appendix_model}

The quantitative trade model we use for our counterfactual simulations is a standard trade model with a positive supply elasticity. It falls into the more general class of models that \citet{AllenArkolakisTakahashi:20} define as the universal gravity framework.
There are $N$ countries, denoted by the subscript $i$ or $j$. Goods are produced by combining labor, which is immobile across countries but mobile within countries, with intermediate inputs. Each country produces a different good that is used for final consumption or as an intermediate good. The production function is a Cobb-Douglas function with constant returns to scale, as in the model of \citet{EatonKortum:02}, where $\zeta$ denotes the share of labor (denoted by $L_i$) in costs and $1 - \zeta$ the share of intermediate inputs (denoted by $M_i$):
\begin{equation*}
    Q_i = A_i L_i^\zeta M_i^{1-\zeta},
\end{equation*}
where $A_i > 0$ is a constant total factor productivity parameter. Intermediate goods consist of the same bundle of goods as those entering final consumption, so the price index for intermediate goods for each firm is the price index for all goods. We denote this price index by $P_i$. Under perfect competition, the price of output at location $i$ is given by
\begin{equation*}
    p_i = {\kappa} (w_i/A_i)^\zeta P_i^{1-\zeta},
\end{equation*}
where ${\kappa} > 0$ is a constant that depends on the parameter $\zeta$, and $w_i$ is the wage rate.

The value of output in the country of origin is $Y_i \equiv p_i Q_i$. Since labor is the only factor of production and profits are zero, the value of all production is distributed to the workers. This gives the usual accounting identity:
\begin{equation*}
    Y_i = p_i Q_i = w_i L_i.
\end{equation*}

In each country there is a representative consumer-worker who supplies labor inelastically and earns the wage $w_i$. This consumer has a constant elasticity of substitution (CES) utility function that aggregates goods from all origins, as in the models of \citet{Armington:69}, \citet{Anderson:79}, and \citet{AndersonvanWincoop:03}. The elasticity of substitution is denoted by $\sigma>1$. The optimization problem of the consumer together with arbitrage in the goods market leads to the well-known result that expenditure on goods from different origins is given by:
\begin{equation*}
X_{ij} = \frac{p_{ij}^{-\epsilon}}{\sum_{k \in S} p_{kj}^{-\theta}} E_j,
\end{equation*}
where expenditure $E_j$ is defined by $E_j \equiv \sum_i X_{ij}$ and $\epsilon \equiv \sigma-1 > 0$ is the trade elasticity.

The representative consumer's indirect utility function is $W_i = w_i/P_i$. As shown by \citet{AllenArkolakisTakahashi:20}, among others, welfare ($W_i$) in this model is proportional to the terms of trade index (defined as the ratio of export price to consumer price) raised to $1+\psi$, where $\psi \equiv (1-\zeta)/\zeta$ is known as the supply elasticity:
\begin{equation*}
    W_i = \frac{w_i}{P_i} = \tilde{\kappa}_i \left(\frac{p_i}{P_i}\right)^{1+\psi},
\end{equation*}
where $\tilde{\kappa}_i > 0$ is a constant.

Exports from country $i$ to country $j$ incur an iceberg trade cost, denoted by $\tau_{ij} \geq 1$. Due to arbitrage in goods markets, this implies that the price paid for imports in country $j$ of the good exported by country $j$ is
\begin{equation*}
    p_{ij} = \tau_{ij} p_i.
\end{equation*}

Equilibrium requires that output markets clear, that is, that prices and quantities adjust so that output $Q_i$ in each country equals the aggregate demand of all countries, including iceberg costs:
\begin{equation*}
    Q_i = \sum_{j=1}^N \tau_{ij} q_{ij},
\end{equation*}
where $q_{ij}$ is the amount of goods that reach the destination country $j$ after deducting the iceberg cost.

Trade deficits are exogenous. We follow \citet{AllenArkolakisTakahashi:20} and assume that for all countries $i$,
\begin{equation*}
    E_i = \Xi \xi_i p_i Q_i,
\end{equation*}
where
\begin{equation*}
    \Xi \equiv \frac{\sum_i p_i Q_i}{\sum_i \xi_i p_i Q_i}.
\end{equation*}
The parameter $\xi_i > 0$ is a constant country-specific parameter. This way of specifying trade deficits ensures that the world's trade deficit is zero.

%This model falls into the more general class of models that \citet{AllenArkolakisTakahashi:20} define as the universal gravity framework. 

We perform counterfactual simulations using the exact hat algebra method by \citep{DekleEatonKortum:08}. %There are several algorithms that can be used. 
To solve for the equilibrium changes, we use the algorithm described by \citet{CamposReggioTimini:24}.

\clearpage
\section{World trade in the quantitative simulation}

\begin{figure}[ht]
    \centering
    \subcaptionbox{Volume}
    {\includegraphics[width=0.45\textwidth]{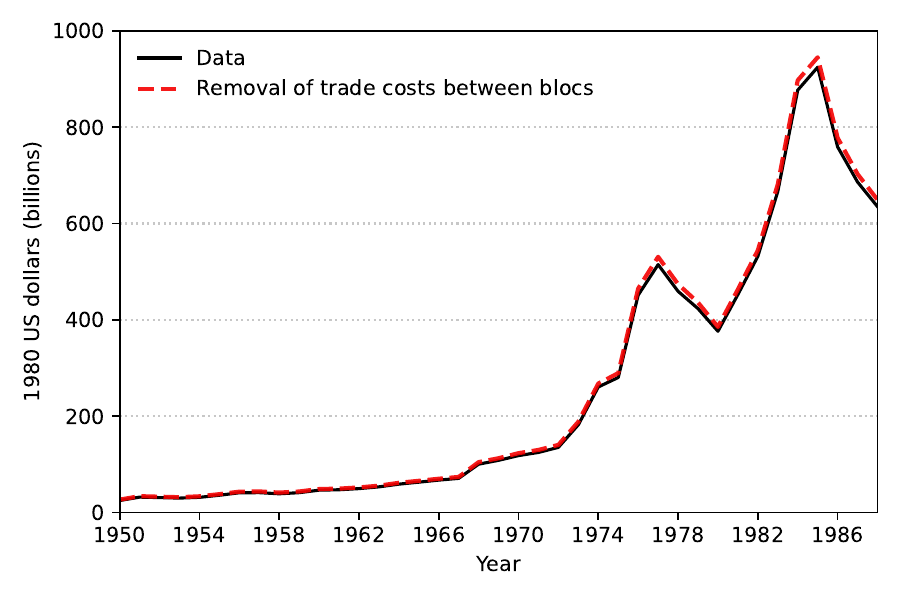}}
    \subcaptionbox{Percent differences}
    {\includegraphics[width=0.45\textwidth]{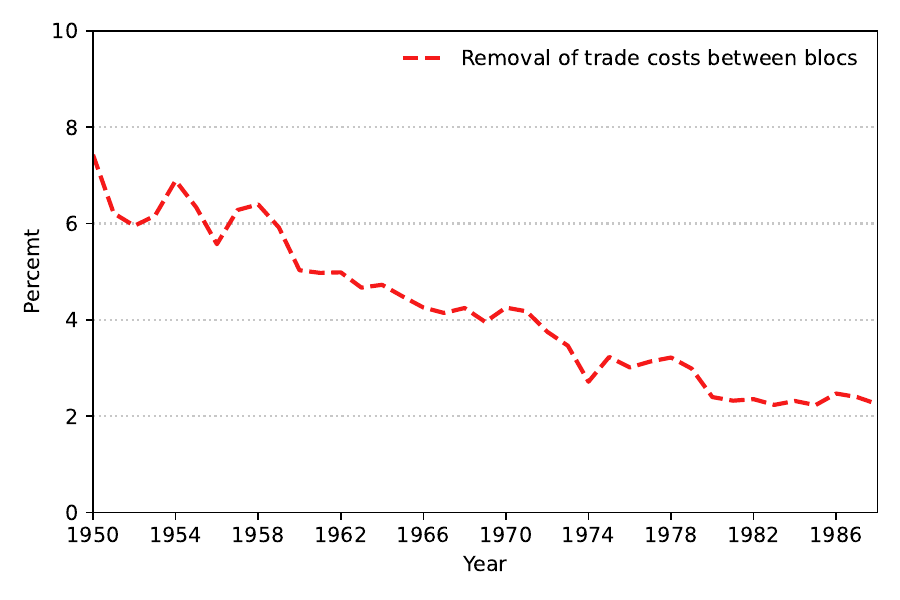}}
    \caption{World trade} \label{fig:counterfactual_total}
    \vspace{11pt}
    \begin{minipage}{0.9\textwidth}
    \footnotesize
    \textbf{Notes}: The figure shows the results of a simulation in which the trade barriers of the Iron Curtain are removed. The solid line in the left panel shows the actual trade volume between East and West. The dashed line shows the counterfactual trade volume. The panel on the right shows the predicted percentage increase in trade volume that would occur if the trade barriers of the Iron Curtain were removed.
    \end{minipage}
\end{figure}

The simulation of Section~\ref{sec:quantification} also yields results for world trade. We report these here. Figure~\ref{fig:counterfactual_total} shows that without the Iron Curtain, world trade would have been about 6\% higher in the 1950s and about 2\% higher in the 1980s.

\end{document}